\documentclass[12pt]{article}
\usepackage{amsfonts}
\input epsf.tex
\font\ninerm=cmr9

\begin{document}

\title{Disordered 2d quasiparticles in class D:\\ Dirac fermions with
  random mass,\\ and dirty superconductors}

\author{M.~Bocquet, D.~Serban,\footnote{Service de Physique
    Th\'eorique, CE-Saclay, F-91191 Gif-Sur-Yvette, France}~ and
  M.R.~Zirnbauer\footnote{Inst. f. Theoretische Physik, Uni K\"oln,
    Z\"ulpicher Str.~77, D-50937 K\"oln, Germany}}

\date{March 15, 2000}
\maketitle

\begin{abstract}
  Disordered noninteracting quasiparticles that are governed by a
  Majorana-type Hamiltonian -- prominent examples are dirty
  superconductors with broken time-reversal and spin-rotation
  symmetry, or the fermionic representation of the 2d Ising model with
  fluctuating bond strengths -- are called class $D$.  In two
  dimensions, weakly disordered systems of this kind may possess a
  metallic phase beyond the insulating phases expected for strong
  disorder.  We show that the 2d metal phase emanates from the free
  Majorana fermion point, in the direction of the RG trajectory of a
  perturbed WZW model.  To establish this result, we develop a
  supersymmetric extension of the method of nonabelian bosonization.
  On the metallic side of the metal-insulator transition, the density
  of states becomes nonvanishing at zero energy, by a mechanism akin
  to dynamical mass generation.  This feature is explored in a model
  of $N$ species of disordered Dirac fermions, via the mapping on a
  nonlinear sigma model, which encapsulates a ${\mathbb Z}_2$
  spin degree of freedom.  We compute the density of states in a finite
  system, and obtain agreement with the random-matrix prediction for
  class $D$, in the ergodic limit.  Vortex disorder, which is a relevant 
  perturbation at the free-fermion point, changes the density of states
  at low energy and suppresses the local ${\mathbb Z}_2$ degree of 
  freedom, thereby leading to a different symmetry class, $BD$.  
\end{abstract}

\section{Introduction}
\label{sec:intro}

As is well-known, the two-dimensional Ising model has a magnetic phase
transition, the critical behavior of which is governed by the
relativistic field theory of a massless Majorana fermion or, on
squaring the partition function, a massless Dirac fermion.  A theme of
some debate over the last fifteen years has been the effect of
disorder on this phase transition.  Disorder, when introduced in the
form of spatial inhomogeneities in the bond strengths, is known to add
a random mass term to the Majorana Lagrangian.  Perturbative
renormalization group calculations by Dotsenko and Dotsenko \cite{DD}
showed that randomness in the mass is a marginally irrelevant
perturbation leading to logarithmic corrections to the pure Ising
critical behavior.  In particular, the singularity in the specific
heat of the Ising model persists (in a weakened form) in the presence
of disorder.  This conclusion was confirmed by Shankar \cite{shankar}
and Ludwig \cite{ludwig}, who also calculated the effect of disorder
on the moments of the spin-spin correlation function.

The picture emerging from this work is that the 2d Ising model with
weakly disordered bond strengths undergoes a phase transition
controlled by the pure Ising fixed point, and the only effect the
disorder has are logarithms.  Although this picture came to be widely
accepted, a constant challenge has been the work of Ziegler
\cite{ziegler90,ziegler96,ziegler98}, who claims that the
thermodynamic singularities of the Ising model are rounded off by the
disorder.  Specifically, he argues that the vanishing density of
states (DoS) of the relativistic fermion at zero energy becomes finite
as a result of nonperturbative effects.  To motivate this scenario, he
refers to a model of disordered Dirac fermions on the lattice, for
which the DoS at $E = 0$ can be proved to be strictly positive.

Ziegler's field-theoretic calculation imitates the standard treatment
of disordered metals, and is based on a Hubbard-Stratonovich
transformation introducing a composite field $Q$, followed by a
saddle-point approximation.  Such a strategy is in principle not
unreasonable.  Indeed, in the case of disordered metals the effective
theory for $Q \sim \psi\bar\psi$, where $\psi$ is the basic electron
field, not only allows the systematic study of quantum interference
corrections to metallic behavior, but also captures the
nonperturbative physics of localization in low dimension.  By analogy,
one expects that in the present case, too, an advantage might be
gained by transforming from Majorana fields $\psi$ to composite fields
$Q \sim \psi\psi$.

An important lesson learned from disordered metals is that, when using
the field-theoretic formulation, we must exercise particular care to
correctly implement the symmetries of the disordered Hamiltonian.  For
example, any inaccuracy in the treatment of time-reversal invariance
(when present) will spoil the weak localization effects due to the
cooperon mode.  What, then, are the symmetries of the disordered
Majorana theory?  The situation is most easily explained if we switch
from two real (Majorana) fermions to the equivalent representation by
one complex (Dirac) fermion, and is as follows.  The first-quantized
Hamiltonian for a two-dimensional Dirac fermion with random mass $m(x)$,
\begin{equation}
  H = \pmatrix{m(x) &-i\partial_1 - \partial_2 \cr
    -i\partial_1 + \partial_2 &-m(x)\cr}
  \;, \label{hamilt}
\end{equation}
has a symmetry of the ``particle-hole'' type:
\begin{equation}
  H = - \sigma_1 H^{\rm T} \sigma_1 \;,
  \qquad \sigma_1 = \pmatrix{0 &1\cr 1 &0\cr} \;.
  \label{phsym}
\end{equation}
According to the general classification of disordered single-particle
systems \cite{az,rss}, this symmetry places the Hamiltonian
(\ref{hamilt}) in class $D$.\footnote{The equation (\ref{phsym}) fixes
  an orthogonal Lie algebra in even dimension, ${\rm so}(2N)$, which
  is denoted by $D_N$ in Cartan's table.  Hence the name ``$D$''.}
{}From the analysis of Ref.~\cite{rss}, one then infers that the
supersymmetric field-theory representation for random-mass Dirac
fermions has a global invariance under the orthosymplectic Lie
supergroup ${\rm OSp}(2n|2n)$,\footnote{Throughout this paper,
  supergroups such as ${\rm GL}(n|n)$ and ${\rm OSp}(2n|2n)$ are
  understood to be defined over the complex number field, ${\mathbb
    C}$, unless specified otherwise.  In places where this fact is of
  particular importance, we will switch to the notation ${\rm
    GL}_{\mathbb C}(n|n)$ and ${\rm OSp}_{\mathbb C}(2n|2n)$.} if the
energy vanishes and the disorder average of a product of $n$ Green
functions is to be calculated.  Actually, the existence of this
orthosymplectic symmetry was first pointed out by D.~Bernard
\cite{bernard} in his Carg\`ese lectures on the application of
conformal field-theory techniques to two-dimensional disordered
systems.  (Random Dirac fermions have also been treated by CFT
techniques in \cite{mcw}.)

It is clear that the existence of such a symmetry will have an
important bearing on the low-energy physics.  For one thing, it was
shown in \cite{az} that the Gaussian random-matrix ensemble for class
$D$ has a {\it positive} density of states at zero energy.  (This can
be understood from the fact \cite{az} that the eigenvalues of a
Gaussian random matrix in class $D$ behave as a noninteracting gas of
harmonically confined fermions on the half-line, with Neumann boundary
conditions at the origin.)  For another, the density of states enjoys
the status of an ``order parameter'' in the field-theoretic formalism.
It is well known that, when the condensation of an order parameter is
associated with a spontaneous breaking of global symmetries, there
appear massless modes due to Goldstone's theorem.  In the present
context, the saddle-point value of the $Q$-field (the order parameter)
breaks ${\rm OSp}(2n|2n)$ symmetry, and Goldstone's theorem hence
forces the existence of a supersymmetric ${\rm OSp}(2n|2n)$ multiplet
of Goldstone modes.  (As usual, these modes have a physical
interpretation as diffusion-like modes.  They are characteristic of
class $D$, and were called the ``spin-singlet $D$-type diffuson'' in
\cite{az}.)  It is a perplexing fact that these Goldstone modes appear
nowhere in Ziegler's work.  Nonetheless, they exist, and because they
do, the fate of the system in the thermodynamic limit cannot be
decided on the basis of a plain saddle-point analysis, but is
determined by the notoriously subtle problem of interacting Goldstone
modes in two dimensions.  Solving this problem requires the use of the
renormalization group.  Thus we are led to ask: once we have augmented
Ziegler's approach by incorporating the orthosymplectic symmetry of
the disordered Majorana theory, what is the prediction for the local
density of states at $E = 0$? Does it vanish in the thermodynamic
limit, or does it not?

Further motivation for the present paper comes from several issues
beyond the density-of-states controversy.  According to the general
scheme of \cite{az}, another realization of symmetry class $D$ is by
the low-energy quasiparticles of disordered superconductors with
broken time-reversal and spin-rotation invariance.  Physically, such a
situation may occur in spin-triplet superconductors, or in
spin-singlet superconductors with spin-orbit scattering, when
time-reversal symmetry is broken spontaneously or by a magnetic field.
Alternatively, both time-reversal and spin-rotation invariance can be
broken by randomly adding classical Heisenberg impurity spins.  With
neither the electric charge nor the spin of a single quasiparticle
being conserved in class $D$, the only constant of the motion is the
energy.  Hence, quasiparticle diffusion and localization in a
superconductor of class $D$ has to be probed via {\it thermal}
transport (or transport of energy).

The qualitative physics of class-$D$ quasiparticles was the subject of
a recent paper by Senthil and Fisher \cite{sfD}.  These authors drew a
schematic phase diagram for the two-dimensional case in particular.
There are three phases, namely thermal insulator, thermal quantum Hall
fluid, and thermal metal, with three distinct phase boundaries, which
meet in a multicritical point, ${\cal M}^*$.  The structure of the
phase diagram is determined on very basic grounds, by localization of
all states for strong disorder (giving the insulating phases), the
renormalization group flow of a weakly coupled nonlinear sigma model
(giving the metal), and the topological distinction between the
insulator and the quantum Hall fluid with edge currents.

In a more speculative attempt, Senthil and Fisher went on to try and
match the phases of disordered 2d quasiparticles in class $D$ with the
phase diagram of the 2d random-bond Ising model.  Such an
identification is prompted by the representation of the 2d Ising model
by a Majorana fermion.  However, there appeared to be a difficulty:
the {\it three} stable phases of symmetry class $D$ in two dimensions
seem to have only {\it two} counterparts, namely the ferromagnet and
the paramagnet, in the Ising model. Where is the missing third phase?
To resolve this puzzle, Senthil and Fisher suggested two alternative
scenarios.  In brief, the first one identifies the multicritical point
${\cal M}^*$ with a point on the Nishimori line \cite{nishimori},
while the second one interprets the missing phase as a spin glass
phase.  (The latter, however, is thought to exist only at zero
temperature in ${\rm d} = 2$.)  Neither scenario looks convincing,
which will motivate us to take a fresh look at the puzzle.

The latest contribution to the subject was made by Read and Green
\cite{rg}, who expounded the fact that another physical realization of
the massless Majorana theory exists (in mean-field approximation) at
the transition between paired quantum Hall states, and in chiral
$p$-wave superconductors at the transition between the topologically
distinct phases of weak and strong pairing.  Concerning the role of
disorder, they suggested to extend the definition of class $D$ so as
to include vortices.  Moreover, they argued that randomly placed
vortices are a strongly relevant perturbation at the free-fermion
point.  The question, however, exactly what the renormalization group
flows to, was not answered conclusively.  Based on symmetry grounds
only, Read and Green wrote down a nonlinear sigma model which differs
from the one we obtain for class $D$ in that the target space in the
fermion-fermion (FF) sector, ${\rm O}(2n)/{\rm U}(n)$, is replaced by
its connected component ${\rm SO}(2n)/{\rm U}(n)$.  Thus we may ask:
what happened to the ${\mathbb Z}_2$ degree of freedom that
distinguishes between ${\rm O}(2n)$ and ${\rm SO}(2n)$?  Giving an
answer to this subtle and yet pertinent question provides the final
motivation for writing the present paper.

In view of the length of the paper, we now summarize our main results.

\subsection{Overview}
\label{sec:overview}

For generality and better perspective, we study a family of models
with $N \ge 1$ species of Dirac fermions, replacing the random mass
$m(x)$ for a single species by a mass matrix $M_{kl}(x)$ for $N$
species.  A perturbative renormalization group calculation shows that
in addition to the random-mass coupling, a few other couplings are
generated by the RG flow for $N > 1$.  By extending the initial
formulation of the model, we incorporate the most important one of
these.  We then use standard technology to approximate the disordered
Dirac theory by an effective action for nonlinear fields $Q : {\mathbb
  R}^2 \to {\bf X}$.  As expected from \cite{rss}, the target space
${\bf X}$ is a Riemannian symmetric superspace of type $C{\rm I}|D{\rm
  III}$.  Its bosonic base (or ``body'') is a product $M_{\rm B}
\times M_{\rm F}$ where $M_{\rm B} = {\rm Sp}(2n, {\mathbb R}) / {\rm
  U}(n)$ is a noncompact symmetric space of type $C{\rm I}$, and
$M_{\rm F} = {\rm O} (2n) / {\rm U}(n)$ is a compact symmetric space
of type $D{\rm III}$.  Note that by ${\rm O}(2n)$ we do mean the {\it
  full} orthogonal group, which consists of {\it two} connected 
components (the orthogonal matrices with determinant $+1$ or $-1$).
Thus the present target space is distinguished by the striking feature
of having two {\it disjoint} components, and there exists the
possibility, not previously encountered in this context, of
forming ${\mathbb Z}_2$ domain walls in the $Q$-field theory.

The effective action for the nonlinear field $Q$ is the logarithm of a
superdeterminant, resulting from integration over the Dirac fields.
One now wants to expand the effective action in gradients of $Q$, to
produce a nonlinear sigma model.  It turns out that performing this
expansion requires a certain amount of care -- the mathematical
subtlety involved is known by the name of chiral anomaly.  To do the
expansion correctly, we have to resort to a variant of nonabelian
bosonization, which is the celebrated statement \cite{witten} that the
free theory of $n$ Majorana fermions has an equivalent representation
by a level-one ${\rm O}(n)$ Wess-Zumino-Novikov-Witten (WZW) model.
Supersymmetry extends this to an equivalence between $2n+2n$ Majorana 
fields (fermions and bosonic ghosts), and a level-one WZW model of fields 
taking values in a Riemannian symmetric superspace of type $C|D$, based 
on ${\cal M}_{\rm B} \times {\cal M}_{\rm F}$ where ${\cal M}_{\rm B} = 
{\rm Sp}(2n,{\mathbb C}) / {\rm Sp}(2n)$ and ${\cal M}_{\rm F} = {\rm
  O}(2n)$.

Using a supersymmetric extension of the bosonization rules for the
chiral densities $\psi_\pm \bar\psi_\mp$, we are then able to compute
the gradient expansion of the effective action.  The Lagrangian of the
$C{\rm I}|D{\rm III}$ 
nonlinear sigma model thus obtained contains a topological term, or
winding number term, with topological coupling $\theta$.  Such a term
is permitted by symmetry, since the massless Dirac Hamiltonian, $H_0$,
depends on the choice of orientation of ${\mathbb R}^2$ or, in other
words, reversing the orientation by a parity transformation, say by
exchanging the two coordinates $x_1$ and $x_2$ of ${\mathbb R}^2$,
takes $H_0$ into an inequivalent Hamiltonian.  (Note that, quite
generally, Dirac operators exist on {\it spin manifolds}, which are
manifolds carrying a spin structure and thus an orientation.)  The
topological term is trivial for $n = 1$, but nontrivial for $n \ge 2$,
since
  $$
  \Pi_2 \left( {\rm O}(2n) / {\rm U}(n) \right) = \left\{
    \begin{array}{lll}
      0\quad &{\rm for} &n = 1 \;, \cr
      {\mathbb Z}\quad &{\rm for} &n \ge 2 \;.
    \end{array} \right.
  $$
By reduction of the multi-valued action of the $C|D$ WZW model,
we show that the topological angle has the value $\theta = N\pi$.

The passage from random-mass Dirac fermions to the nonlinear sigma
model is under good control, {\it i.e.}~the terms omitted are small,
for $N \gg 1$.  The limit of a large number $N$ of species is of
course unrealistic.  Fortunately, we can also control the case $N =
2$, if the kinetic energy is {\it anisotropic}, with the first species
being more mobile in the $x_1$ direction and the second more mobile in
the $x_2$ direction.  A closely related situation is relevant for the
application of our results to disordered $d$-wave superconductors with
a subcritical concentration of localized impurity spins \cite{gbs}.
For $N \gg 1$, or $N = 2$ with large anisotropy, the nonlinear sigma
model is at weak coupling, and the one-loop beta function predicts
renormalization group flow to a Gaussian fixed point describing a
perfect metal.  The local density of states at $E = 0$ in this case diverges
logarithmically in the thermodynamic limit.  We also compute the
density of states for a finite system in the ergodic regime, and
obtain agreement with the random-matrix prediction of \cite{az}.

In contrast, for $N = 1$ the mapping on the nonlinear sigma model is
far from being controlled.  Even if we trust the mapping, the model is
strongly coupled, and no safe statements can be easily made from it.  
We have to concede that our method fails in that case, and the problem 
is better analysed through its original formulation in terms of Dirac
fields.  Our conclusion thus is the one commonly accepted: random 
mass is a marginally irrelevant perturbation, and the theory in the
infrared flows to the free-fermion point, where the density of states
at zero energy vanishes.

There now seems to be a conflict, but only superficially so, with 
Ziegler's rigorous proof of a positive lower bound for the density of 
states. The apparent discrepancy is resolved by observing that 
Ziegler works with a naive lattice discretization of the Dirac operator,
thereby imposing an additional lattice symmetry on his model, as a
result of which the Hilbert space decomposes into {\it two decoupled
  sectors}.  In each of these, the Hamiltonian can be shown to be a
pi-flux model (details of the argument will be published in a separate
comment), which belongs to the time-reversal invariant Wigner-Dyson
class $A{\rm I}$ and, in fact, is known \cite{ff,hl} to have a finite
density of states at zero energy.  Hence Ziegler's rigorous result,
albeit correct, is a statement about a different symmetry class, and
does not falsify our results.  We may safely ignore his proof in all
that follows.

There is another point of possible contention that deserves to be made
clear.  Recall that our large-$N$ saddle-point analysis yields a
target manifold consisting of {\it two} connected components.  Thus
there is a ${\mathbb Z}_2$ degree of freedom in the effective theory,
and we anticipate the existence of ${\mathbb Z}_2$ domains and domain
walls, across which the nonlinear sigma model field jumps from one
connected component of the target manifold to the other.  (Note that
in our computation of the one-loop beta function of the $C{\rm I} |
D{\rm III}$ nonlinear sigma model, the nonperturbative effect of these
domain walls is neglected.) The same ${\mathbb Z}_2$ degree of
freedom arises for small $N$, when the method of nonabelian
bosonization is used.  Indeed, nonabelian bosonization according to
Witten \cite{witten} transforms $n$ Majorana fermions into a WZW model
over the {\it full} orthogonal group ${\rm O}(n)$, not just its
identity component ${\rm SO}(n)$. The necessity to work with ${\rm O}
(n)$, which consists of two connected components, is readily seen by
looking at the special case $n = 1$: a single Majorana fermion is not
an empty theory, as would be the case with the trivial group ${\rm
  SO}(1)$, but is equivalent to a theory of local ${\rm O}(1) \equiv
{\mathbb Z}_2$ degrees of freedom, namely the 2d Ising model.  In a
sense, then, passing from random-mass fermions to the effective
description by a nonlinear sigma model, reintroduces the local Ising
degrees of freedom underlying the free fermion.

Having established the general formalism and its validity, we turn to
a specific application: disordered non-chiral $d$-wave superconductors
with broken time-reversal and spin-rotation symmetry.  Following a
standard procedure \cite{fradkin,ntw,sl}, we linearize the dispersion
relation around four low-energy points (or ``nodes'') in wave vector
space, and then add generic disorder (in the form of a random scalar
potential, random complex order parameter, and random spin-orbit
scattering) to place the model in class $D$.  The number of species
per node is $N = 2$ due to spin.  By drawing on our previous results,
we map the low-energy physics of the quasiparticles on a nonlinear
sigma model.  Doing this for every node separately, we obtain kinetic
energy terms that are spatially anisotropic, and a topological angle
which is positive for two of the nodes and negative for the other two.
The inclusion of quasiparticle scattering between the nodes merges the
four anisotropic models into a single isotropic theory.  In the last
step, all the topological terms cancel, as is required for a
non-chiral superconductor invariant under reflections of space.

Next, we recall the 2d phase diagram proposed by Senthil and Fisher
\cite{sfD}, and address the puzzle why the {\it two} stable phases of
the 2d random-bond Ising model seem to have {\it three} counterparts
in class $D$.  To resolve this mismatch, one needs to understand the
nature of the multicritical point ${\cal M}^*$ and the renormalization
group flow in its vicinity.  Our resolution of the puzzle is very
simple: the ``multicritical'' point ${\cal M}^*$, where the three
phases of class $D$ meet, is to be identified with the free-fermion
point!  The evidence in favor of such an identification is twofold.
Firstly, by an argument involving nonabelian bosonization, we show
that the free-fermion point sits right on the boundary of the 2d metal
phase.  Secondly, we argue that (barring vortex disorder) the thermal
metal phase of class $D$ can only be realized in the enlarged
parameter space available for more than one species ($N > 1$), and is
absent in the fundamental case $N = 1$.  This makes it natural to
invert some of the flows assumed in \cite{sfD} and arrive at a
plausible 2d phase diagram, where the paramagnetic and ferromagnetic
states of the Ising model match the thermal insulator and thermal
quantum Hall fluid phases of class $D$ for $N = 1$, and there is no
redundant third phase.

Another suggestion by Senthil and Fisher \cite{sfD} says that, from
the perspective of the nonlinear sigma model, the existence of the
insulating phases, which are distinguished by a quantized thermal Hall
conductance, can be attributed to the presence of the topological term
in the action functional.  One might then think that this very term is
what is driving the phase transition from the thermal insulator to the
thermal quantum Hall fluid.  That scenario is widely accepted for a
close cousin, namely Pruisken's nonlinear sigma model \cite{pruisken}
of the integer quantum Hall effect.  However, as was stated earlier,
the topological term of the $C{\rm I}|D{\rm III}$ nonlinear sigma
model is {\it trivial} for $n = 1$.  Since the phase transition occurs
for all $n$, including $n = 1$, it is hard to see how the topological
term could be its driving agent.  A better scenario is to attribute
the phase transition to the ${\mathbb Z}_2$ degree of freedom of the
$C{\rm I}|D{\rm III}$ nonlinear sigma model.

Finally, we delve into the quasiparticle physics of chiral $p$-wave
superconductors, to shed more light on the effect of vortex disorder
in those systems.  Read and Green \cite{rg} have argued, both
intuitively and formally, that randomly placed vortices are a relevant
perturbation at the free-fermion point of a superconductor with
(mean-field) order parameter symmetry $p_{x+iy}$.  Unlike the
perturbation by a random mass matrix, which can become relevant only
in the extended parameter space available for $N \ge 2$, vortex
disorder turns out to be relevant even in the fundamental case of
spinless particles $(N = 1)$.  Read and Green further propose that
vortices drive the superconductor from the free-fermion point into a
thermal metal phase, the effective field theory for which {\it lacks},
as it stands, the ${\mathbb Z}_2$ (or Ising spin) degree of freedom of
the nonlinear sigma model we obtain for random-mass fermions.  We
believe that their proposal is correct, and in order to provide
supporting evidence for it, we outline an indirect argument, passing
through a variant of the Chalker-Coddington network model with local
${\rm O}(1)$ invariance.  (We are unable at present to give a direct
field-theoretic proof, as we do not know how to incorporate vortex
disorder into the nonabelian bosonization scheme.)

What we are learning, then, is that vortex disorder exerts a drastic
influence on the effective field theory: while there exists a local
${\mathbb Z}_2$ degree of freedom (and domain walls in it) when
vortices are absent, this degree of freedom is suppressed when
vortices are inserted.  Thus the cases with and without vortices map
on {\it different} field theories.  Actually, there must exist a whole
one-parameter family of such theories, as the suppression proceeds
continuously.  Motivated by random-matrix limits, we propose to refer
by the (hybrid) name $BD$ to the generic symmetry class including a
variable amount of vortex disorder.  Class $D$ is then viewed as a
subclass of the generic class $BD$.  Of course, the field-theoretic
distinction between classes does not fail to leave its imprint on the
quasiparticle physics.  We expect that when all quasiparticle states
are localized, {\it i.e.}~in the thermal insulator and quantum Hall
fluid phases, the local density of states at zero energy vanishes in
the absence of vortices (class $D$), but becomes finite when an
extensive number of vortices is inserted (class $BD$).

In contrast, the physical distinction between systems with and
without vortices is of relatively minor consequence in the
metallic regime of extended states.  There, and for the case of a
finite system and energies $E$ below the Thouless energy $E_{\rm Th}$,
the nonlinear sigma model can be evaluated in zero-mode approximation.
For the total density of states in the two cases at hand,
we obtain:
  $$
  \rho_D(E) = \nu + {\sin(2\pi\nu E) \over 2\pi E} \;, \qquad
  \rho_{BD}(E) = \nu + {\textstyle{1\over 2}}\delta(E) \;,
  $$
where the scale of variation is set by $\nu$, the inverse of the level
spacing for energies much greater than the mean level spacing (but
still less than $E_{\rm Th}$).  The first expression results on
integrating over the full target manifold (with ${\mathbb Z}_2$
present), the second from the reduced target (with ${\mathbb Z}_2$
completely suppressed).  These results coincide and, as we shall see, 
not accidentally so, with the large-$N$ limits of the density of states
for the Haar random-matrix ensembles on ${\rm SO}(2N)$ and 
${\rm O}(2N)$, respectively.

The material of the paper is arranged as follows.  In Sections
\ref{sec:hamilt}--\ref{sec:susy}, we introduce a Hamiltonian for $N$
species of Dirac fermions with random mass, and set up a
field-theoretic representation of its Green functions by a
supersymmetric Gaussian functional integral.  Renormalization of the
disorder-averaged field theory generates a total of four marginal
perturbations, the flow equations for which are worked out in Section
\ref{sec:marginal}.  Guided by this, we settle for a specific choice
of Lagrangian and proceed to map it on the $C{\rm I}|D{\rm III}$
nonlinear sigma model, in the limit $N \gg 1$.  The composite field
$Q$ is introduced and subjected to a saddle-point approximation in
Section \ref{sec:Qfield}, while the structure of the saddle-point
manifold is elucidated in Section \ref{sec:saddles}.  Section
\ref{sec:gradient} is the longest of the paper.  There, we expand the
effective action for $Q$ in gradients, using a supersymmetric
extension of the method of nonabelian bosonization, which is developed
in two major subsections.  In Section \ref{sec:DOS}, we renormalize
the nonlinear sigma model (neglecting the ${\mathbb Z}_2$ degree of
freedom) and show that its coupling flows to zero, resulting in a
perfect metal with infinite thermal conductivity.  We compute the
density of states for a finite system, as well as in the thermodynamic
limit.  In Section \ref{sec:sc} we clarify the relation between the
model considered and disordered $d$-wave superconductors.  Section
\ref{sec:phases} assembles various arguments that focus on the role of
the free-fermion point inside class $D$, and culminate in the proposal
of a local phase diagram.  Finally, in Section \ref{sec:vortices} we
comment on the role of vortices and the significant modification they
cause in the nonlinear sigma model.

To avoid confusion, we emphasize that the present paper will always be
concerned with the physics of {\it Majorana fermions}.  Nevertheless,
the words ``Dirac fermions'' or ``Dirac theory'' will be frequently
encountered, the reason being that we carry out the disorder average
by using the supersymmetry method, and a single Majorana fermion does
not have a bosonic analog.  The standard trick to get around this
problem is to ``square the partition function'', making two Majorana
fermions out of one, and then combine the two to form a Dirac fermion.
The latter can be augmented by a bosonic $b$-$c$ ghost system to make
the theory supersymmetric and cancel all vacuum graphs.

\section{Dirac Hamiltonian with random mass}
\label{sec:hamilt}

The first-quantized Hamiltonian $H$ for $N$ species of Dirac fermions
with random mass in two dimensions is written as
\begin{equation}
  H = \pmatrix{M(x) &-2i\partial\cr -2i\bar\partial &-M^{\rm T}(x)\cr} \;.
  \label{Hamilt}
\end{equation}
Here $M = (M_{kl})$ is an $N \times N$ mass matrix, and we use the
convention $\partial = {1\over 2} (\partial_1 -i \partial_2)$, $\bar
\partial = {1\over 2}(\partial_1+i\partial_2)$.  Aside from being
Hermitian ($H^\dagger = H$), this Hamiltonian enjoys an important
symmetry property:
  $$
  H = -\sigma_1 H^ {\rm T} \sigma_1 \;,
  $$
which will be called particle-hole symmetry, or the symmetry of class
$D$.  The superscript ${\rm T}$ denotes joint transposition in
particle-hole and position space (note $\partial^{\rm T} = -
\partial$).  The particle-hole symmetry of $H$ is dictated by the
Lagrangian for Majorana fermions,
  $$
  L_M = i\bar\psi_l \partial \bar\psi_l +
    i \psi_l \bar\partial \psi_l + \bar\psi_k M_{kl} \psi_l \;.
  $$
(Summation over repeated indices is understood.)  An immediate
consequence of the particle-hole symmetry is that the eigenvalues of
$H$ occur in pairs with opposite sign $\pm E$.  Hence, the density of
states is symmetric with respect to the point $E = 0$.  By the same
token, there exists a relation between retarded and advanced Green
functions $G^\pm(E) = (E \pm i0 - H)^{-1}$:
  $$
  G^-(E) = - \sigma_1 G^+(-E)^{\rm T} \sigma_1 \;.
  $$
When the mass matrix is a constant multiple of unity, $M(x) = m
\times 1_N$, one easily computes the local density of states:
\begin{eqnarray*}
  \nu(E) &=& ({\rm area})^{-1} \pi^{-1}
  {\rm Im}\;{\rm Tr} \left( E-i0-H \right)^{-1}\\
  &=& \frac{2N}{\pi} \;\lim_{\epsilon \to 0+}\; {\rm Im}
  \int \frac{d^2k}{(2\pi)^2} \;\frac{E-i \epsilon}
  {(E-i \epsilon)^2-k^2-m^2} \\
  &=& \frac{N |E|}{2\pi}\; \theta (E^2-m^2) \;,
\end{eqnarray*}
which vanishes linearly at $E = 0$ in the limit of zero mass.

In what follows, we take the entries of the mass matrix to be Gaussian
distributed random variables with zero mean and second moments given
by
\begin{equation}
  \big\langle M_{ij}(x) M_{kl}(y) \big\rangle =
  (2g_M / N) \, \delta_{il} \, \delta_{jk} \, \delta(x-y) \;.
  \label{disorder}
\end{equation}
The density of states on average over the fluctuating mass matrix can
be obtained by computing the average Green function.  This will be
done by an adaptation of Efetov's supersymmetry method \cite{efetov}.
Although that method is in principle quite standard, we will see that
its application to the present situation features some peculiarities
which are well worth explaining.

\section{Supersymmetric integral representation}
\label{sec:susy}

We are going to employ a supersymmetric integral representation to set
up the calculation of the disorder averaged Green functions.  Although
we concentrate on the case of a single Green function ($n = 1$) for
notational simplicity, the generalization to arbitrary $n > 1$ will be
immediate \cite{rss}.  We introduce $2N$ supermultiplets $\phi_k$ and
$\bar\phi_l$ $(k,l = 1, ..., N)$, each containing two fermionic
components $\psi_{\pm}$ and two bosonic superpartners $b$ and
$c$.\footnote{We use the conformal field theory convention of denoting
  antiholomorphic fields by an overbar.  To avoid confusion where it
  might otherwise arise, complex conjugation is denoted by an asterisk
  $*$ instead of the overbar.}  Each supermultiplet is arranged in the
form
  $$
  \phi = \pmatrix{\psi_-\cr \psi_+\cr  c \cr  b \cr} \;, \qquad
  \bar\phi = \pmatrix{\bar\psi_-\cr \bar\psi_+\cr \bar c \cr \bar b \cr}\;,
  $$
where the index $l = 1,...,N$ was omitted.  The orthosymplectic transpose 
$\phi^{\rm t}$ of $\phi$ is defined by
\begin{equation}
  \phi^{\rm t} \equiv ( \psi_+ , \psi_- ,  b  , - c  ) \;,
  \label{transpose}
\end{equation}
and $\bar\phi^{\rm t}$ (defined similarly) is the orthosymplectic
transpose of $\bar\phi$.  Note that this choice of transpose
determines an inner product which is {\it skew}:
\begin{equation}
  \bar\phi^{\rm t} \phi^{\vphantom{t}} = \bar\psi_{+}\psi_{-} +
  \bar\psi_{-}\psi_{+} +  \bar b c -  \bar c b =
  - \phi^{\rm t} \bar\phi^{\vphantom{t}} \;.
  \label{skew}
\end{equation}
Moreover, the orthosymplectic transpose on $\phi$ defines a compatible
transpose of $4 \times 4$ supermatrices $T$ by the requirement $(T
\phi)^{\rm t} \equiv \phi^{\rm t} T^{\rm t}$.

The field-theory Lagrangian for $N$ species is taken to be
\begin{equation}
  L_0 = \bar\phi_l^{\rm t} \partial \bar\phi_l^{\vphantom{t}}
  + \phi_l^{\rm t} \bar\partial \phi_l^{\vphantom{t}}
  + i \bar\phi_k^{\rm t} M_{kl}^{\vphantom{t}} \phi_l^{\vphantom{t}}
  - i E \bar\phi_l^{\rm t} \Sigma_3 \phi_l^{\vphantom{t}} \;,
  \label{L0}
\end{equation}
where
  $$
  \Sigma_3 = 1_{\rm susy} \otimes \sigma_3 = {\rm diag}(+1,-1,+1,-1) \;,
  $$
and summation over repeated $k,l$ indices is understood.  If $j$ and
$\bar j$ are source fields, the Green functions of the Dirac
Hamiltonian (\ref{Hamilt}) are generated by the functional
  $$
  {\cal Z}[j] = \Big\langle \int {\cal D}\bar\phi \, {\cal D}\phi
  \, \exp - \int d^2x \left(L_0 +\bar\phi^{\rm t}j + \phi^{\rm t}
    \bar j \right) \Big\rangle \;,
  $$
as is easily verified from the fact that $L_0$ is the quadratic form
constructed by sandwiching the operator
  $$
  1_{\rm susy} \otimes i(H-E) = 1_{\rm susy} \otimes
  \pmatrix{iM-iE &2\partial\cr 2\bar\partial &-iM^{\rm T}-iE\cr}
  $$
between
  $$
  (\bar\psi_+,\bar b ,\psi_+, b ) \quad {\rm and} \quad
  \pmatrix{\psi_-\cr  c \cr \bar\psi_-\cr \bar c \cr} \;,
  $$
and integrating by parts to symmetrize the terms with derivatives.  In
order for the Gaussian functional integral over $\phi, \bar\phi$ to
make sense, the bosonic fields must be related to each other by
complex conjugation $*$:
\begin{equation}
   b_l^* = \bar c_l^{\vphantom{*}} \;, \quad
   c_l^* = \bar b_l^{\vphantom{*}} \;.
  \label{conjug}
\end{equation}
Given this convention, the functional integral is purely oscillatory
for $E \in {\mathbb R}$, and converges for energies in the upper half
plane (${\rm Im}E > 0$).  As stated in the introduction, the
Lagrangian $L_0$ in the limit $E \to 0$ acquires an invariance under
global transformations,
\begin{eqnarray*}
  &&\phi_l(x) \mapsto T \cdot \phi_l(x) \;, \quad
  \bar\phi_l(x) \mapsto T \cdot \bar\phi_l(x) \;, \\
  &&\phi_l^{\rm t}(x) \mapsto \phi_l^{\rm t}(x) \cdot T^{\rm t} \;, \quad
  \bar\phi_l^{\rm t}(x) \mapsto \bar\phi_l^{\rm t}(x) \cdot T^{\rm t} \;,
\end{eqnarray*}
for elements $T \in {\rm OSp}(2|2)$, {\it i.e.}~if $T$ obeys $T^{\rm
  t}T = 1$, with $T^{\rm t}$ being the orthosymplectic supermatrix
transpose defined above. 

On performing the disorder average specified by (\ref{disorder}), the
Lagrangian for the case $E = 0$ becomes
\begin{equation}
  L_1 = \bar\phi_l^{\rm t} \partial \bar\phi_l^{\vphantom{t}}
  + \phi_l^{\rm t} \bar\partial \phi_l^{\vphantom{t}} +
  (g_M/N) \Phi^{(1)} \;, \qquad \Phi^{(1)} = \bar\phi_k^{\rm t}
  \phi_l^{\vphantom{t}} \bar\phi_l^{\rm t} \phi_k^{\vphantom{t}} \;.
  \label{Lag1}
\end{equation}
By simple power counting, the operator $\Phi^{(1)}$ is a perturbation
of the massless Dirac theory which is marginal in the renormalization
group sense.  $\Phi^{(1)}$ may be marginally relevant or marginally
irrelevant.  The question of which is the case, is decided by
calculating the short-distance expansion of the operator product of
$\Phi^{(1)}$ with itself.

\section{Marginal perturbations}
\label{sec:marginal}

It turns out that the operator product expansion of $\Phi^{(1)}$ with
itself generates three additional operators:
\begin{eqnarray*}
  \Phi^{(2)} &=& \bar\phi_k^{\rm t} \bar\phi_l^{\vphantom{t}}
    \phi_l^{\rm t} \phi_k^{\vphantom{t}} \;, \\
  \Phi^{(3)} &=& \bar\phi_k^{\rm t} \phi_k^{\vphantom{t}}
    \bar\phi_l^{\rm t} \phi_l^{\vphantom{t}} \;, \\
  \Phi^{(4)} &=& \bar\phi_k^{\rm t} \phi_l^{\vphantom{t}}
    \bar\phi_k^{\rm t} \phi_l^{\vphantom{t}} \;,
\end{eqnarray*}
for $N > 1$.  Each of these is marginal and invariant under ${\rm
  OSp}(2|2)$, and the set $\Phi^ {(1)}, ..., \Phi^{(4)}$ exhausts the
set of marginal perturbations permitted by that symmetry.  Note that
the situation simplifies for $N = 1$, where $\Phi^{(1)} \equiv
\Phi^{(3)} \equiv \Phi^{(4)}$, and $\Phi^{(2)} \equiv 0$ from
(\ref{skew}).

Given that the operators $\Phi^{(\alpha)}$ $(\alpha = 2, 3, 4)$ are
generated by the renormalization group flow anyway, the natural
procedure is to include them in the bare theory.  Our goal thus is to
renormalize the extended Lagrangian
\begin{equation}
  L_2 = \bar\phi_l^{\rm t} \partial \bar\phi_l^{\vphantom{t}}
  + \phi_l^{\rm t} \bar\partial \phi_l^{\vphantom{t}}
  + N^{-1} \sum_{\alpha=1}^4 g_\alpha \Phi^{(\alpha)} \;.
  \label{Lag2}
\end{equation}
To do so, we use the fact \cite{zamoRG,opebeta} that, if the leading
short-distance singularity of the operator product expansion (OPE) is
  $$
  \Phi^{(\alpha)}(x) \Phi^{(\beta)}(x^\prime) = |x-x^\prime|^{-2}
  \sum_\gamma C_\gamma^{\alpha\beta} \Phi^{(\gamma)}(x^\prime) + ... \;,
  $$
the one-loop beta functions determining the RG flow with increasing
cutoff length scale $\ell$, are given by
  $$
  {2\over\pi}\dot g_\gamma\equiv {dg_\gamma\over d\ln\ell} = -{\pi\over N}
  \sum_{\alpha\beta} C_\gamma^{\alpha\beta} g_\alpha g_\beta \;.
  $$
The expansion of the operator product $\Phi^{(\alpha)}(x) \Phi^{(\beta)}
(x^\prime)$ is completely determined by the OPEs for holomorphic $(z =
x_1 + ix_2)$ and antiholomorphic $(z^* = x_1-ix_2)$ fields.  These are
\begin{equation}
  \phi_k^a(z) \phi_l^b(w) \sim {1\over 2\pi} \,{\delta_{kl} \,
    \tau^{ab} \over z-w} \;,
  \qquad \bar\phi_k^a(z^*) \bar\phi_l^b(w^*) \sim {1\over 2\pi}
  {\delta_{kl} \, \tau^{ab} \over z^* - w^*} \;,
  \label{opedirac}
\end{equation}
where
  $$
  \tau = \sigma_1 \otimes E_{\rm FF} + i\sigma_2 \otimes E_{\rm BB}
  = \pmatrix{0 &1 &0 &0\cr 1 &0 &0 &0\cr 0 &0 &0 &1\cr 0 &0 &-1 &0\cr}
  $$
determines the orthosymplectic structure.  Using these formulas, a
lengthy but straightforward calculation yields
\begin{eqnarray}
  \dot g_1 &=& g_1 g_2 - N^{-1} \left(
    g_1 g_2 + g_1 g_4
    - g_2 g_3 + g_3 g_4 \right) \;, \nonumber \\
  \dot g_2 &=& {\textstyle{1 \over 2}} (g_1^2 + g_2^2)
  + N^{-1} \left( g_1 g_3 + g_1 g_4 - g_2^2
    - g_3 g_4 \right) \;, \nonumber \\
  \dot g_3 &=& - (g_1+g_2)g_3 - N^{-1} \left(
    g_1^2 - g_1 g_2 + g_1 g_4 + g_2
    g_3 + g_3^2 + g_3 g_4 \right) \;, \nonumber \\
  \dot g_4 &=& - N^{-1} \left( g_1 g_2 + g_1 g_3
    + g_2 g_3 + g_4^2 \right) \;.
  \label{flow}
\end{eqnarray}
For $N = 1$ these equations simplify greatly.  In that case,
$\Phi^{(2)} \equiv 0$, so the coupling $g_2$ must be dropped, and
since $\Phi^{(1)} = \Phi^{(3)} = \Phi^{(4)}$, the only coupling to
renormalize is $g_M \equiv g_1 + g_3 + g_4$.  By linearly combining the
above equations, we get
\begin{equation}
  {dg_M \over d\ln\ell} = - {2 g_M^2 \over \pi} \;,
  \label{irrelevant}
\end{equation}
which reproduces the well-known result \cite{DD} that random mass
is a marginally irrelevant perturbation (for $N = 1$).

Let us quickly review how to deduce from this result the behavior of
the local density of states $\nu(E,g_M)$ at small energies $E$ and
weak coupling $g_M$.  We start from the formula
  $$
  \nu(E,g_M) = \pi^{-1} {\rm Re} \left\langle (\psi_- \bar\psi_+
    + \bar\psi_- \psi_+)(0) \right\rangle \;,
  $$
where the functional average is defined w.r.t. the Lagrangian
(\ref{Lag1}) for $N = 1$.  The OPE between $\Phi^{(1)}$ and the
operator that couples to $E$, $\Phi^{(0)} \equiv \bar\phi^{\rm t}
\Sigma_3 \phi$, reads $\Phi^{(1)}(x) \Phi^{(0)}(0) = - (2\pi^2)^{-1}
\Phi^{(0)} (0) / |x|^2 + ...$.  It then follows, by the same principle
we just used for the coupling $g_M$, that the operator $\Phi^{(0)}$
has scaling dimension $\gamma(g_M) = 1 - g_M/\pi + {\cal O}(g_M^2)$,
and the RG flow equation for $E$ is $dE / d\ln\ell = (2-\gamma(g_M))
E$.  By making in the functional integral an RG transformation which
changes the cutoff from $\ell_0$ to $\ell$, we obtain
  $$
  \nu\left( E(\ell_0) , g_M(\ell_0) ; \ell_0 \right) =
  \nu\left( E(\ell) , g_M(\ell) ; \ell \right) \exp \, -
  \int_{\ell_0}^\ell \gamma(g_M(\ell^\prime)) d\ln\ell^\prime \;.
  $$
We know from (\ref{irrelevant}) that the coupling $g_M(\ell)$ flows to
zero with increasing $\ell$.  We also know the density of states of the
pure system to be independent of the cutoff: $\nu(E,0;\ell) = |E|/2\pi$.
Hence, by integrating the flow equations for $g_M$ and $E$ up to the
infrared cutoff, which for small enough energies is given by the system
size $L$, we obtain
  $$
  \nu(E,g_M;\ell_0) = {|E| \over 2\pi} \left(
  1 + (2 g_M/\pi) \ln (L/\ell_0) \right) \;.
  $$
We see that, for $N = 1$, the density of states at $E = 0$ remains
zero, and randomness in the mass only changes the slope of $\nu \sim
|E|$ by a logarithm.

For $N > 1$ the flow equations for the couplings $g_1,...,g_4$ are not
very transparent, and the precise form of the RG flow remains unclear
in general.  However, for $N \gg 1$ the equations reduce to
  $$
  \dot g_1 \pm \dot g_2 = \pm {\textstyle{1 \over 2}}
  (g_1 \pm g_2 )^2 \;, \quad  \dot g_3 = -
  (g_1+g_2)g_3 \;, \quad \dot g_4 = 0 \;.
  $$
Clearly, the parameters $g_3$ and $g_4$, if initially set to zero,
remain zero under the flow.  The relevant couplings are $g_1$ and
$g_2$, which are seen to be attracted to the line $g = g_1 = g_2$,
with $g$ satisfying the equation
\begin{equation}
  {dg \over d\ln\ell} = \beta(g) = + {2 g^2 \over \pi} \;.
  \label{gflow}
\end{equation}
We observe that the sign on the right-hand side has been reversed as
compared to $N = 1$, and the coupling $g$ is now {\it relevant}.

These considerations motivate us to adopt the modified Lagrangian
\begin{equation}
  L_3 = \bar\phi_l^{\rm t} \partial \bar\phi_l^{\vphantom{t}}
  + \phi_l^{\rm t} \bar\partial \phi_l^{\vphantom{t}}
  + (g/N) \left( \Phi^{(1)} + \Phi^{(2)} \right) \;.
  \label{modLang}
\end{equation}
At the level of the Dirac Hamiltonian (\ref{Hamilt}), the inclusion
of the operator $\Phi^{(2)}$ corresponds to adding a term
  $$
  H \to H + \pmatrix{0 &\Delta\cr \Delta^\dagger &0\cr} \;,
  $$
where $\Delta$ is a skew-symmetric $N \times N$ matrix whose entries
are Gaussian distributed random variables with zero mean and second
moments
  $$
  \big\langle \Delta_{ij}(x) \bar\Delta_{kl}(y) \big\rangle
  = (2g / N) (\delta_{ik}\delta_{jl} - \delta_{il}\delta_{jk})
  \,\delta(x-y) \;.
  $$
Such a term is permitted by the fundamental particle-hole symmetry
(\ref{phsym}) of class $D$.  Note that for $N = 1$, skew-symmetry
forces $\Delta$ to vanish identically.  (Equivalently, $\Phi^{(2)} =
0$ for $N = 1$, and $L_3$ coincides with the original Lagrangian.)
For the application to disordered superconductors (Section
\ref{sec:sc}), we put $N = 2$.  The indices $i,j,k,l$ then label the
spin degrees of freedom of the electron, while the random functions
$M(x)$ and $\Delta(x)$ acquire a physical meaning as random scalar
potential, random spin-orbit scattering, and the random part of
the superconducting order parameter (or, to be precise, linear
combinations thereof).

Before embarking on our project of analysing $L_3$, we wish to mention
three other special choices of the couplings, where we can easily
understand the nature of the RG flow.  The first case is $g_1 = g_3 =
0$, $N > 2$.  By tracing the couplings back to a {\it Hermitian}
random Hamiltonian, we find the constraint $g_1 \pm g_4 \ge 0$.  Thus,
setting $g_1 = 0$ forces $g_4 = 0$.  Inspection shows that the random
Hamiltonian in this limit has a higher degree of symmetry:
\begin{eqnarray*}
  H &=& - (\sigma_1 \otimes 1_N) H^{\rm T} (\sigma_1 \otimes 1_N) \\
  &=& + (\sigma_2 \otimes 1_N) H^{\rm T} (\sigma_2 \otimes 1_N) \;,
\end{eqnarray*}
where the second equality is a kind of time-reversal invariance,
placing $H$ in class $D{\rm III}$ \cite{az}.  The only coupling that
remains is $g_2$, which flows according to $\dot g_2 \propto (N-2)
g_2^2$ and thus is relevant for $N > 2$. The methods developed in the
present paper are readily adapted to deal with this case.

The second special case is obtained by setting $N = 2$ in the previous
one.  The underlying random Hamiltonian then is of the form
  $$
  H = \pmatrix{0 &-2i\partial + a \sigma_2\cr
    -2i\bar\partial + \bar a \sigma_2 &0 \cr} \;.
  $$
Because this commutes with ${\rm diag}(\sigma_2,\sigma_2)$, the
Hilbert space decomposes into two sectors invariant under the action
of $H$.  (For $N > 2$, the Hilbert space does not decompose.)  The
Hamiltonian restricted to either sector is a Dirac fermion in a random
abelian vector potential $\pm a(x)$, which is class $A{\rm III}$
\cite{az,rss}, and explains \cite{lfsg} why the single coupling $g_2$
is exactly marginal for $N = 2$.  The extra time-reversal symmetry
here does no more than connect the two restrictions of $H$ by a Lie
algebra homomorphism.  Apart from causing every energy eigenvalue to
be doubly degenerate, this is of no consequence, and the symmetry class
is $A{\rm III}$ instead of $D{\rm III}$.

There exists a third special case where the flow equations close on
a single coupling.  This is $g_2 = g_3 = 0 = g_1 + g_4$ with $g_1 > 0$,
and again $N = 2$.  Here the underlying random Hamiltonian is of the
form
  $$
  H=\pmatrix{V\sigma_2 &-2i\partial\cr -2i\bar\partial &V\sigma_2\cr}\;.
  $$
By a simple conjugation $H^\prime = U H U^{-1}$, this can be
transformed into $H^\prime = {\rm diag}(H_+ , H_-)$ where
  $$
  H_{\pm} = \pmatrix{\pm V &-2i\partial\cr -2i\bar\partial &\pm V\cr}
  $$
are two copies of a Dirac fermion with random scalar potential $V(x)$.
In the transformed basis, each copy satisfies
  $$
  H_\pm = \sigma_2 H_\pm^{\rm T} \sigma_2 \;,
  $$
which is the defining equation of the ``symplectic'' Wigner-Dyson
class $A{\rm II}$.  The extra particle-hole symmetry (class $D$) here
does no more than relate the first copy to the second copy by a Lie
algebra homomorphism.  It is therefore redundant, and the symmetry
class remains $A{\rm II}$.

By numerically solving the flow equations (\ref{flow}), we find that
the line $g_2 = g_3 = 0 = g_1 + g_4$ has a large basin of attraction
for $N = 2$.  In particular, it attracts the initial condition 
$g_1 = g_2$, $g_3 = g_4 = 0$ in that case. 

According to \cite{rss}, the classes $D$, $D{\rm III}$ and $A{\rm II}$
exhaust the set of symmetry classes whose field-theory representation
is acted upon by the present form of ${\rm OSp}(2n|2n)$ (preserving a
symplectic structure for bosons and an orthogonal structure for
fermions).  We therefore expect no further accidental closures to
occur.

\section{Composite field and diagonal saddles}
\label{sec:Qfield}

As we saw, for $N \gg 1$ the coupling $g = g_1 = g_2$ is certain to
increase under renormalization, leading to a strong-coupling problem
in the infrared.  To handle this problem, it is mandatory to switch to
a dual formulation in terms of a Hubbard-Stratonovich field $Q \sim
\phi_l^{\vphantom{t}} \bar\phi_l^{\rm t} + \bar\phi_l^{\vphantom{t}}
\phi_l^{\rm t}$.

The composite field $Q$ is brought on stage by doing two Gaussian
integrals in a row.  To prepare these steps, we reorganize the
Lagrangian (\ref{modLang}) as follows:
  $$
  L_3 = \bar\phi_l^{\rm t} \partial \bar\phi_l^{\vphantom{t}}
  + \phi_l^{\rm t} \bar\partial \phi_l^{\vphantom{t}} + \frac{g}{2N}
  \, {\rm STr} \left( \phi_k^{\vphantom{t}} \bar\phi_k^{\rm t} +
    \bar\phi_k^{\vphantom{t}} \phi_k^{\rm t} \right)
  \left( \phi_l^{\vphantom{t}} \bar\phi_l^{\rm t} +
    \bar\phi_l^{\vphantom{t}} \phi_l^{\rm t} \right) \;.
  $$
The quartic interaction can now be decoupled in a first step by
introducing an auxiliary $4 \times 4$ supermatrix field $Q$:
\begin{eqnarray}
  L_4 &=& \bar\phi_l^{\rm t} \partial \bar\phi_l^{\vphantom{t}}
  + \phi_l^{\rm t} \bar\partial \phi_l^{\vphantom{t}} + {\rm STr} \, Q
  (\phi_l^{\vphantom{t}} \bar\phi_l^{\rm t} + \bar\phi_l^{\vphantom{t}}
  \phi_l^{\rm t}) - \frac{N}{2g} \, {\rm STr} \, Q^2  \nonumber \\
  &=& (\bar\phi_l^{\rm t}\,\phi_l^{\rm t}) \pmatrix{Q &\partial\cr
    \bar\partial &Q\cr} \pmatrix{\phi_l\cr \bar\phi_l\cr} -
  \frac{N}{2g}\,{\rm STr} \, Q^2 \;. \label{decouple}
\end{eqnarray}
In the second step, one does the Gaussian integral over the Dirac
field, resulting in the action functional
\begin{equation}
  S_5[Q] = - {N \over 2g}\int d^2x \,{\rm STr}\,Q^2 + {N \over 2}\,
  {\bf STr}\ln\pmatrix{\Sigma_3 Q &\partial\cr \bar\partial &Q\Sigma_3\cr}\;,
  \label{action}
\end{equation}
where ${\bf STr}$ combines the operations of taking the supertrace
${\rm STr}$ and integrating over position space.  The factors
$\Sigma_3$ under the logarithm appear after the transformation
$\bar\phi_l \mapsto \Sigma_3 \bar\phi_l$ and $\bar\phi^{\rm t} \mapsto
\bar \phi^{\rm t} \Sigma_3$, correcting for the fact that $\bar\phi$
is not the complex conjugate of $\phi^{\rm t}$; see the definition
(\ref{transpose}) and the relations (\ref{conjug}).  The above result
is for $E = 0$.  To restore the dependence on energy, we need to shift
$Q \to Q - (iE/2) \Sigma_3$ under the argument of the logarithm.  Note
that all manipulations done so far were exact.

As a result of the transformation to $Q$, the parameter $N$ now
appears as a factor in the exponent, while $Q$ itself is ignorant of
the number of species.  For large $N$, this suggests treating the $Q$
integral in saddle-point approximation, which is what we are going to
do next.  (Note, however, that the saddle-point approximation is
uncontrolled for $N = 1$.) The saddle-point equation reads
  $$
  Q(x)/g = {\textstyle{1 \over 2}} \left\langle x \left|
      (Q-\partial Q^{-1}\bar\partial)^{-1} \right| x \right\rangle
  + (\partial \leftrightarrow \bar\partial) \;.
  $$
We look for a spatially homogeneous solution of the form $Q(x) =
\mu \Sigma_3$.  The saddle-point equation then reduces to
  $$
  g^{-1} = \int {d^2 k \over (2\pi)^2} \, (\mu^2 + k^2/4)^{-1} \;.
  $$
Cutting off the integral in the ultraviolet by $|k| < 1/\ell_0$ yields
the equation
  $$
  \pi / g = \ln \left( 1 + (2\mu\ell_0)^{-2} \right) \;,
  $$
and by inversion,
\begin{equation}
  \mu = (2\ell_0)^{-1} / \sqrt{{\rm e}^{\pi/g} - 1} \;.
  \label{genmass}
\end{equation}
For weak disorder $(g \ll 1)$, this is well approximated by $\mu
\approx (2\ell_0)^{-1} {\rm e}^{-\pi/2g}$.  From formula (\ref{gflow})
we then infer that $\mu$ obeys the renormalization group equation
  $$
  \left({d\over d\ln\ell}+\beta(g){d\over dg}\right)\mu(\ell,g)=0\;,
  $$
and thus has the meaning of a {\it dynamically generated mass}.  In
other words, the mass scale $\mu$ remains invariant:
\begin{equation}
  \mu(\ell_0,g(\ell_0)) = \mu(\ell,g(\ell)) \;,
  \label{dynmass}
\end{equation}
under an RG transformation $\ell_0 \mapsto \ell$.  This means that the
Dirac field undergoes the phenomenon of dimensional transmutation:
while the bilinear $\bar\phi_l^{\rm t}\phi_l^{\vphantom{\rm t}}$ has
dimension one at the free-fermion point, its dimension approaches zero
on large scales.  At the same time, $Q \sim \phi_l^{\vphantom{\rm t}}
\bar\phi_l^{\rm t} + \bar\phi_l^{\vphantom{\rm t}}\phi_l^{\rm t}$
acquires a nonvanishing expectation value.

In principle, all diagonal matrices $Q$ with entries $\pm\mu$ are
candidate solutions of the saddle-point equation.  It may happen,
however, that some of these do not lie on the integration domain, or
cannot be reached by analytic continuation from that domain.
Therefore, in order to be able to select the proper solutions, we have
to be more explicit about which integration domain to choose for the
superfields $\phi_l$, $\bar\phi_l$ and $Q$.

\section{Saddle-point manifold}
\label{sec:saddles}

The issue of how to choose correctly the integration domain for the
superfields $\phi,\bar\phi$ and for the supermatrix $Q$ was carefully
addressed in \cite{rss}.  Drawing on (but not repeating) the detailed
analysis of that reference, we argue as follows.  Let ${\rm Im}E > 0$
for definiteness.  Then, in order to have convergent integrals over
the bosonic ghosts of the Dirac version of the theory, we must impose
the conditions (\ref{conjug}) or, equivalently, $\phi_{a,{\rm
    B}}^\dagger = \bar\phi_{a,{\rm B}}^{\rm t}\sigma_3$.  Note that
there is no need to require such a relation among the Grassmann
fields, and we do not impose any such requirement.

Next one easily verifies that the bilinear $A \equiv \phi_l^{
  \vphantom{t}} \bar\phi^{\rm t}_l +\bar\phi_l^{\vphantom{t}}
\phi_l^{\rm t}$ in the Lagrangian $L_4$ is odd under the
orthosymplectic transpose: $A^{\rm t} = - A$, so $A \in {\rm osp}
(2|2)$, by definition of the orthosymplectic Lie algebra.  Via its
coupling to $A$ in $L_4$, the supermatrix $Q$ inherits the same
property:
  $$
  Q = - Q^{\rm t} \in {\rm osp}(2|2) \;.
  $$
Now, in the process of decoupling the interaction $\Phi^{(1)} + \Phi
^{(2)}$, severe convergence problems arise in the boson-boson (BB)
block denoted by $Q_{\rm BB}$.  However, close inspection shows that
the integrals over $Q_{\rm BB}$ can be arranged to exist (without
ruining the convergence of the integrals over the bosonic Dirac ghosts
$b,c$) by choosing the following parametrization:
  $$
  Q_{\rm BB} =  Y + \gamma \,{\rm e}^X \Sigma_3 \, {\rm e}^{-X} \;,
  $$
where $X, Y \in {\rm osp}(2|2)$ are odd resp. even with respect to
conjugation by $\Sigma_3$:
  $$
  \Sigma_3 X \Sigma_3 = - X \;, \quad \Sigma_3 Y \Sigma_3 = Y \;,
  $$
and the even (or bosonic) parts $X_0, Y_0$ of $X, Y$ are subject to
  $$
  X_0^{\vphantom{\dagger}} = X_0^\dagger \;, \quad
  Y_0^{\vphantom{\dagger}} = - Y_0^\dagger \;.
  $$
The parameter $\gamma > 0$ is in principle arbitrary but, anticipating
its saddle-point value, we set it to $\gamma = \mu$.  This deals with
the BB sector.  The situation in the FF sector is more benign.  There,
convergence of the integrals over $Q_{\rm FF}$ is simply achieved by
requiring $Q_{\rm FF} = Q_{\rm FF}^\dagger$.

Now we can compare the above parametrization for $Q$ to the diagonal
solutions of the saddle-point equation.  The structure of the solution
in the BB block is uniquely fixed by the convergence requirements to
be
  $$
  Q_{0,{\rm BB}} = \mu \sigma_3 \;.
  $$
Other diagonal solutions cannot be reached by deformation of the
integration manifold without crossing some singularities of the
integrand.

In contrast, in the FF sector analyticity provides no selection rule
on the saddle points; there, the integrand does not possess any poles,
but only has zeroes as a function of $Q_{\rm FF}$, so that the path of
integration can be analytically continued to cross any saddle point in
that sector.  However, in the weak-coupling (or large $N$) limit, some
of the saddles contribute in a negligible manner.  The dominant
contributions come from those that minimize the super-dimension of the
transverse manifold. This extremality condition is fulfilled when the
positive and negative eigenvalues of $Q_{0,{\rm FF}}$ are equal in
number.  In the present case, this criterion leaves two possibilities:
  $$
  Q_{0,{\rm FF}} = \pm \mu \sigma_3 \;.
  $$
In summary, we retain as dominant (diagonal) saddle points:
\begin{equation}
  Q_0 = \mu \;(\pm E_{\rm FF} + E_{\rm BB}) \otimes \sigma_3 \;.
  \label{sdp}
\end{equation}

Recall now the existence of a global $G \equiv {\rm OSp}(2|2)$
symmetry (at $E = 0$).  The symmetry group acts on the supermatrix $Q$
by conjugation: $Q(x) \mapsto T Q(x) T^{-1}$ ($T \in G$).  Clearly,
such transformations leave the action functional $S_5[Q]$ in
(\ref{action}) invariant.  As a result, the saddles of $S_5[Q]$ are
degenerate: given any solution $Q_0$ of the saddle-point equation, we
get an {\it orbit} of solutions by acting on $Q_0$ with the symmetry
group $G$.  Because the stability condition $h Q_0 h^{-1} = Q_0$
divides out a subgroup $H = {\rm GL}(1|1)$, called the stabilizer of
$Q_0$, the orbit of $G$ on $Q_0$ is a coset space $G/H = {\rm
  OSp}(2|2) / {\rm GL}(1|1)$.  Notice that this is the orbit of the
{\it complex} symmetry group.

To carry out the saddle-point integral, we need to restrict the
bosonic degrees of freedom of $G/H$ to a {\it real} submanifold.  We
now briefly describe this submanifold, setting the Grassmann variables
temporarily to zero.  The bosonic part of the complex supergroup ${\rm
  OSp}(2|2)$ is ${\rm O}(2,{\mathbb C}) \times {\rm Sp}(2,{\mathbb
  C})$.  In the BB sector, the symmetry group is ${\rm Sp}(2,{\mathbb
  C})$, which arises as the group of transformations $T$,
  $$
  \pmatrix{b\cr c\cr} \mapsto T \cdot \pmatrix{b\cr c\cr} \;, \qquad
  \pmatrix{\bar b\cr\bar c\cr}\mapsto T\cdot\pmatrix{\bar b\cr\bar c\cr} \;,
  $$
leaving invariant the symplectic form
  $$
  \bar b  c -  \bar c  b = (\bar b , \bar c)
  \pmatrix{0 &1\cr -1 &0\cr} \pmatrix{b\cr c\cr} \;.
  $$
Those symplectic transformations of $(b , c)$ and $(\bar b , \bar c)$
that preserve the reality conditions (\ref{conjug}), form a real
subgroup ${\rm Sp}(2,{\mathbb R})$.  Division by the stabilizer,
which is isomorphic to ${\rm U}(1) \in H$, then yields the BB manifold
$M_{\rm B} = {\rm Sp}(2,{\mathbb R})/{\rm U}(1) \simeq {\rm H}^2$.
This is a two-hyperboloid. (For general $n$, we get ${\rm Sp}(2n,
{\mathbb R}) / {\rm U}(n)$, which is a noncompact symmetric space of
type $C{\rm I}$.)

In the FF sector, the symmetry group acts by ${\rm O}(2,{\mathbb C})$,
which is understood to be the invariance group of the symmetric form
(or ``orthogonal'' structure)
  $$
  \bar\psi_{+}\psi_{-}+\bar\psi_{-}\psi_{+}=(\bar\psi_{+},\bar\psi_{-})
  \pmatrix{0 &1\cr 1 &0\cr} \pmatrix{\psi_{+}\cr \psi_{-}\cr} \;.
  $$
We have emphasized the crucial fact that we {\it never} impose any
reality conditions on the fermions.  What determines then the real FF
subgroup?  The answer is that reality here enters through the
Hermiticity constraint $Q_{\rm FF} = Q_{\rm FF}^\dagger$ (needed
for convergence), which selects from the complex symmetry group ${\rm
  O}(2,{\mathbb C})$ a real subgroup ${\rm O}(2)$, the usual
orthogonal group in two dimensions.  Dividing by the stabilizer ${\rm
  U}(1)$, we obtain $M_{\rm F} = {\rm O}(2) / {\rm U}(1) \simeq
{\mathbb Z}_2$.  (For general $n$, we get ${\rm O}(2n) / {\rm U}(n)$,
a compact symmetric space of type $D{\rm III}$.)  Thus the base of 
the saddle-point manifold is $M_{\rm B} \times M_{\rm F} = {\rm H}^2
\times {\mathbb Z}_2$. On reinstating the Grassmann variables, we
arrive at a saddle-point supermanifold, ${\bf X}_1$, which is a
Riemannian symmetric superspace of type $C{\rm I}|D{\rm III}$
\cite{rss}.

A distinctive feature of ${\bf X}_1$ is that, instead of being
connected, it consists of {\it two} disjoint pieces.  This fact was
already discovered in \cite{rss}, and we are now going to elaborate
briefly.  The special orthogonal group ${\rm SO}(2)$ acts on two
Majorana fermions $(\xi,\eta)$ as
  $$
  \pmatrix{\xi\cr \eta\cr} \mapsto
  \pmatrix{\cos\theta &-\sin\theta\cr \sin\theta &\cos\theta\cr}
  \pmatrix{\xi\cr \eta\cr} \;,
  $$
while the corresponding ${\rm SO}(2)$ action on the Dirac fermion
$\psi_\pm = \xi \pm i \eta$ is
  $$
  \pmatrix{\psi_+\cr \psi_-\cr} \mapsto
  \pmatrix{{\rm e}^{i\theta} &0\cr 0 &{\rm e}^{-i\theta}\cr}
  \pmatrix{\psi_+\cr \psi_-\cr} \;.
  $$
This does not exhaust the symmetries of the FF sector. The Majorana
fermion is known to possess an additional discrete symmetry, which is
reflection $\xi \leftrightarrow \eta$ (an element of ${\rm O}(2)$ with
determinant minus one), or in Dirac language:
  $$
  \pmatrix{\psi_+\cr \psi_-\cr} \mapsto
  \pmatrix{0 &i\cr -i &0\cr} \pmatrix{\psi_+\cr \psi_-\cr}
  = - \sigma_2 \pmatrix{\psi_+\cr \psi_-\cr} \;.
  $$

Consider now the action of these symmetry transformations on the saddle
point $Q_{0,{\rm FF}} = \sigma_3$ in the FF sector.  The action of
${\rm SO}(2)$ simply fixes the saddle point: ${\rm e}^{i\theta\sigma_3}
\sigma_3 {\rm e}^{-i\theta\sigma_3} = \sigma_3$, and thus is trivial,
while the discrete ${\rm O}(2)$ transformation by $\sigma_2$ reverses
the sign: $\sigma_3 \mapsto \sigma_2 \sigma_3 \sigma_2 = -\sigma_3$.
The latter then explains the origin of ${\mathbb Z}_2$ in the symmetric
superspace ${\bf X}_1$.

In summary, all of the saddle points of $S_5[Q]$ lie on a {\it single}
orbit of the bosonic symmetry group ${\rm O}(2) \times {\rm Sp}(2,
{\mathbb R}) \subset {\rm OSp}(2|2)$, but the orbit is not connected,
as ${\rm O}(2)$ comes in two pieces.  We mention in passing that this
peculiar feature of universality class $D$ also occurs in class $D{\rm
  III}$ \cite{rss}.

Although we have focused on the formalism for a single Green function
$(n = 1)$, all of our considerations readily extend to the case $n \ge
1$.  Supervectors and supermatrices simply become longer or bigger.
The symmetry group inflates to $G = {\rm OSp} (2n|2n)$, the matrix
$\Sigma_3$ tensors to $\Sigma_3 = 1_{\rm susy} \otimes \sigma_3
\otimes 1_n$, and the stabilizer becomes $H = {\rm GL}(n|n)$.  The
order parameter space for a general value of $n$ is denoted by ${\bf
  X}_n$, and has bosonic submanifold
  $$
  M_{\rm B} \times M_{\rm F} =
  \left( {\rm Sp}(2n,{\mathbb R}) / {\rm U}(n) \right)
  \times \left( {\rm O}(2n) / {\rm U}(n) \right) \;.
  $$
Again this consists of two disjoint pieces, corresponding to the
two connected components of ${\rm O}(2n)$ (with determinant plus or
minus one).  In the following section, we shall work with the
extension to general $n$.

As an important corollary to the discovery of a ${\mathbb Z}_2$ degree
of freedom in the saddle-point manifold, we anticipate the existence
of ${\mathbb Z}_2$ domains and domain walls, across which $Q$ jumps
from one connected component of the saddle-point manifold to the
other.  We expect this Ising-like degree of freedom to be very
important for the phenomenology of the insulating phases of class $D$
(Section \ref{sec:phases}), where $Q$ fluctuates strongly.  On the
other hand, when the field is stiff, domain walls are costly in
energy, and therefore they should be of minor relevance in the
metallic limit we shall focus on in Sections
\ref{sec:gradient}--\ref{sec:sc}.

\section{Gradient expansion}
\label{sec:gradient}

To summarize the current state of affairs, the problem at hand has a
global $G = {\rm OSp}(2n|2n)$ symmetry, which is broken to ${\rm GL}
(n|n)$ by making a naive saddle-point approximation.  Whether the
symmetry is truly broken or not, will have to be decided at a later
stage.  (The Mermin-Wagner-Coleman theorem, stating that continuous
symmetries of compact type cannot be broken spontaneously in two
dimensions, does not apply in the present case, as the saddle-point
manifold is a noncompact superspace.  Hence, the question whether
symmetry breaking occurs or not remains open for now.)  Our
saddle-point analysis has identified an order parameter $Q = \mu q$,
where $q \equiv T \Sigma_3 T^{-1}$ takes values in a symmetric
superspace ${\bf X}_n$.  The low-energy configurations of the action
$S_5[Q]$ in (\ref{action}) are given by slowly varying fields
 $$
 q(x) = T(x) \Sigma_3 T(x)^{-1} \quad (T(x) \in G) \;.
 $$
These are the Goldstone modes of the broken $G$ symmetry.  Note that
$q$ lies (on an adjoint orbit of $G$) in ${\rm osp}(2n|2n)$, and
satisfies the constraint $q^2 = 1$.

Our next goal is to derive the low-energy effective action for the
Goldstone modes $q(x)$.  (We shall neglect field fluctuations
transverse to the saddle-point manifold, since these are massive.  We
shall also assume $q(x)$ to be smooth, which means we will ignore
domain walls in the ${\mathbb Z}_2$ degree of freedom.) On general
grounds, the effective action must be of the form
\begin{equation}
  S_{\rm eff}[q] = {-1\over 16 \pi f} \int d^2x \,
  {\rm STr} \, \partial_\mu q \, \partial_\mu q \\
  + {\theta \over 32\pi} \int d^2x \, \epsilon_{\mu\nu} \,
  {\rm STr}\, q \, \partial_\mu q \, \partial_\nu q \;.
  \label{NLsM}
\end{equation}
The two terms displayed are the only ones that contain no more than
two derivatives, respect rotational invariance of position space, and
are compatible with global $G$ symmetry. (Later we will add a
symmetry-breaking term for finite energy $E \not= 0$.)  The second
term is a topological or winding number term.  It arises by pulling
back the closed two-form ${\rm STr}\,q\,{\rm d}q \wedge {\rm d}q$
(called the K\"ahler form) of ${\bf X}_n$, and is nontrivial since
  $$
  \Pi_2({\bf X}_n) = \Pi_2(M_{\rm B} \times M_{\rm F}) =
  \Pi_2 \left( {\rm O}(2n) / {\rm U}(n) \right) =
  \Pi_1 \left( {\rm U}(n) \right) = {\mathbb Z}
  $$
for $n > 1$.  (In contrast, the $n = 1$ winding number $\Pi_2({\rm O}
(2) / {\rm U} (1)) = 0$ is trivial.  It may seem curious that there
exists such a marked difference between $n = 1$ and $n > 1$, as the
topological term is supposed \cite{sfD} to play an important role in
controlling the phase transition between the two insulating phases of
class $D$ in two dimensions.  We will suggest an explanation later,
when we discuss the phase diagram.)  Quantitatively put, for any closed
surface $\Sigma$ we have
  $$
  {1\over 32\pi i}\int_\Sigma d^2x \, \epsilon_{\mu\nu} \,
  {\rm STr} \, q \, \partial_\mu q \, \partial_\nu q \in {\mathbb Z} \;.
  $$
For $n = 2$ the correct normalization factor $(32\pi i)^{-1}$ can be
figured out by direct calculation, using ${\rm O}(4) / {\rm U}(2)
\simeq {\rm S}^2 \times {\mathbb Z}_2$.  The generalization to $n > 2$
is dictated by the connection with the multi-valued action $\Gamma[M]$
in equation (\ref{multival}) below.  We note that the topological term
is odd under parity $x_1 \leftrightarrow x_2$.  Nevertheless, its
presence is permitted, as a choice of pure Dirac Hamiltonian $H_0 =
\sigma_1 p_1 + \sigma_2 p_2$, as compared to $H_0 = \sigma_2 p_1 +
\sigma_1 p_2$, implies a choice of orientation of position space.

Our task now is to work out what the values of the couplings $f$ and
$\theta$ are.  This is easily done for $f$, by straightforward gradient
expansion of
\begin{equation}
  S_5[q] = {N \over 2} \ln {\bf SDet} \pmatrix{\mu \Sigma_3 q
    &\partial\cr \bar\partial &\mu q\Sigma_3\cr} \;. \label{eff_act}
\end{equation}
On the other hand, the value of the topological angle $\theta$ is a
subtle issue.  This is {\it not} easily found by gradient expansion.
The reason is that topological excitations such as instantons, whose
topological charges are what is counted by the winding number term,
necessarily involve {\it large} variations of the field $q$, thereby
defying simple considerations based on Taylor expansion of the action
functional. One viable option would be to evaluate both $S_{\rm eff}
[q]$ and $S_5[q]$ on a well-chosen configuration $q(x)$, say an
instanton solution (such solutions exist for $n > 1$) and equate the
two answers to determine $\theta$.  Unfortunately, the evaluation of
$S_5[q]$ on an instanton is neither easy nor rewarding (which is to
say that it does not lead to any significant payoff beyond solving the
technical problem at hand). A powerful standard trick for computing
determinants of Dirac operators is to take the logarithmic derivative
with respect to a parameter $s$, and integrate over $s$ at the end.
Such a strategy is doomed to fail here, as differentiating a winding
number necessarily produces zero (and, moreover, topologically
distinct sectors cannot be continuously connected with one another).

The most elegant and painless scheme for extracting the low-energy
effective action from $S_5[q]$ proceeds via the method of nonabelian
bosonization.  Following a celebrated paper by Witten \cite{witten},
this method has become a standard tool for dealing with perturbations
of the free fermion theory.  Unfortunately, in the supersymmetric
context of Dirac fermions augmented by a $b$-$c$ ghost system, the
method is less established.  Therefore, in the subsections that are
appended below, we will make a digression to explain the needed
extension of nonabelian bosonization, using the language of functional
integrals.  For now, we describe how the method is used to achieve our
goal of expanding $S_5[q]$ in gradients.

We start by undoing the integration over the fields $\phi,\bar\phi$:
  $$
  {\rm e}^{-S_5[q]/N} = \int {\cal D}\phi \, {\cal D}\bar\phi \,
  \exp \, - \int d^2x \, \left( \bar\phi^{\rm t} \partial \bar\phi
    + \phi^{\rm t} \bar\partial \phi + \mu \bar\phi^{\rm t} q
    \phi + \mu \phi^{\rm t} q \bar\phi \right) \;.
  $$
The notation is the same as before, except that we have dropped the
summation over the species index $(l = 1,...,N)$ and divided $S_5$ by
$N$ accordingly.  The next step is to bosonize, using the principle of
Bose-Fermi equivalence in two dimensions.  Witten's nonabelian version
asserts that the free theory of $2n$ Majorana or $n$ Dirac fermions
has an equivalent representation by a Wess-Zumino-Novikov-Witten (WZW)
model with target ${\rm O}(2n)$ resp.~${\rm U}(n)$ at level $k = 1$.
As will be shown in Subsections \ref{sec:wzw} and \ref{sec:nonab}
below, this equivalence can be generalized to allow for the presence
of a $b$-$c$ ghost system on the free-fermion side.  The equivalent
``bosonized'' theory turns out to be what might loosely, but only
loosely, be called an ``${\rm OSp} (2n|2n)$'' WZW model.  We will give
the precise definitions later.  Here we simply state that the WZW
field, which we denote by $M$, takes values in a subspace of the
complex supergroup ${\rm OSp}(2n|2n)$, and the action functional of
the WZW model has the usual form,
\begin{equation}
  W[M] = {1\over 16\pi}\int_\Sigma d^2x\,{\rm STr}\, (M^{-1}
  \partial_\mu M)^2 + {i\Gamma[M] \over 24\pi} \;. \label{wznw}
\end{equation}
Following Witten, the multi-valued functional $\Gamma[M]$ is expressed
by assuming some extension $\tilde M$ of $M$ to a 3-ball ${\cal B}$
that has position space $\Sigma$ for its boundary ($\partial{\cal B} =
\Sigma$):
\begin{eqnarray}
  \Gamma[M] &=& \int_{\cal B} {\rm STr} \, (\tilde M^{-1}{\rm d}
  \tilde M)^{\wedge 3} \nonumber \\
  &=& \int_{\cal B} d^3x \, \epsilon_{\mu\nu\lambda} \, {\rm STr} \,
  \tilde M^{-1}\partial_\mu \tilde M\, \tilde M^{-1}\partial_\nu \tilde M
  \, \tilde M^{-1}\partial_\lambda \tilde M\;.\label{multival}
\end{eqnarray}
Of course, in order for this to work we must assume the position space
$\Sigma$ to be a surface without boundary, say a two-sphere or a
two-torus.

The nonabelian bosonization rules tell us to replace the free Dirac
theory plus $b$-$c$ ghost system by the WZW model with action $W[M]$,
and the bilinears $\phi\bar\phi^{\rm t}\Sigma_3$ and $\Sigma_3\bar\phi
\phi^{\rm t}$ by $\ell^{-1} M$ resp.~$\ell^{-1} M^{-1}$. (The factor
$\ell^{-1}$ is a large mass scale which enters for dimensional reasons
and depends on the regularization scheme.)  Doing so, and setting $T =
{\rm e}^X$ with $X = - \Sigma_3 X \Sigma_3$, so that $q \Sigma_3 = T
\Sigma_3 T^{-1} \Sigma_3 = {\rm e}^{2X} = T^2$, we obtain
  $$
  {\rm e}^{-S_5/N} = \int{\cal D}M \, \exp \, \left( - W[M] -
    {\mu \over \ell}\int d^2x\,{\rm STr}(M T^{-2} + T^2 M^{-1})\right)\;.
  $$
Next, we remove the factor $T^2 = {\rm e}^{X}$ from the second term
in the exponent, by changing integration variables from $M$ to $M^\prime
= M T^{-2}$:
  $$
  {\rm e}^{-S_5/N} = \int{\cal D}M^\prime\,\exp\,\left(- W[M^\prime
    T^2]-{\mu\over\ell}\int d^2x\,{\rm STr} (M^\prime+M^{\prime -1})
  \right) \;.
  $$
(Note that, since the adjoint orbit of $T$ on $\Sigma_3$ consists of
two disjoint components, this change of variables can only be valid if
$M$ runs through two components, too.)  $T^2$ now appears in the
argument of the WZW functional $W$, while the other term has become a
plain mass term.

The last step is to argue that for small length $\ell$, or large mass
$\mu$, the WZW field $M^\prime$ fluctuates only weakly around
$M^\prime = 1$.  Indeed, the potential ${\rm STr} (M + M^{-1})$ has an
absolute minimum at $M = 1$, as follows from the definition of the
target space as a Riemannian symmetric superspace, see equation
(\ref{minimum}) below.  The leading approximation then is to set
$M^\prime$ simply to unity, which yields
  $$
  S_5[q]\Big|_{q = T\Sigma_3 T^{-1}} = N \, W[T^2] \;.
  $$
Calculating the first correction due to fluctuations of $M^\prime$
around 1, one finds a term with four gradients, multiplied by the
square of the length scale $\sqrt{\ell/\mu}$.  This four-gradient term
becomes small, and the approximation of setting $M^\prime = 1$ is
therefore justified, on length scales larger than $\sqrt{\ell/\mu}$.
Note that if the saddle-point approximation is applied after
renormalization has brought the initially small value of $g$ to about
unity, the two length scales $\ell$ and $\mu^{-1} \approx 2\ell
\sqrt{g/\pi}$ are parametrically the same.  It is then this length
scale $\mu^{-1}\sim \ell$ which will set the short-distance cutoff
of the nonlinear sigma model.

By inserting $M = T^2 = {\rm e}^{2X}$ with $X = - \Sigma_3 X\Sigma_3$
into the first term of $W[M]$ and comparing with $S_{\rm eff}[q]$ for
$q = T \Sigma_3 T^{-1} = {\rm e}^{2X} \Sigma_3$, we readily infer $f =
1 / N$.  This result for $f$ is easy to verify by direct gradient
expansion of $S_5$, without passing through the WZW model.  A more
accurate calculation, taking into account the finite value of
$\mu/\ell$, gives
\begin{equation}
  f = {1 \over N (1 - e^{-\pi/g})} \;. \label{coupling}
\end{equation}

Now we tackle the delicate task of calculating the coupling $\theta$.
Since we expect the topological term of the nonlinear sigma model to
arise from the topological term $\Gamma[M]$ of the WZW model, we
substitute $M = T^2$ into $\Gamma[M]$.  Doing so, we naively get zero
-- the technical reason is that ${\bf X}_n$ does not support such a
term -- but only naively so.  The point to observe is that writing
$\Gamma[M]$ in the form (\ref{multival}) requires a {\it smooth}
extension of $M$ to the ball ${\cal B}$.  If we insist on $M \Sigma_3
= T^2 \Sigma_3 = q$ taking values in ${\bf X}_n$, such an extension
does not exist, by the topological obstruction $\Pi_2( {\bf X}_n)
\not= 0$ (for $n > 1$).

We can avoid the obstruction by allowing $q$ to vary over a {\it
  larger} set, say the complex group ${\rm OSp}(2n|2n)$.  Then, if $0
\le s \le 1$ is a radial coordinate for the ball ${\cal B}$, we can
take the extension to be
  $$
  \tilde M(x,s) = T(x) \exp\left(\pm is\pi\Sigma_3/2\right)
  T(x)^{-1} (\mp i\Sigma_3) \;.
  $$
It is seen that for $s = 1$ we have $\tilde M(x,1) = T(x) \Sigma_3 T(x)
^{-1} \Sigma_3 = q(x)\Sigma_3$, while for $s = 0$ we get $\tilde M(x,0)
= \mp i \Sigma_3$, independent of $x$.  By inserting this extension
into the expression (\ref{multival}) for $\Gamma[M]$, and converting
the integral over ${\cal B}$ into an integral over $\Sigma = \partial
{\cal B}$ using Stokes' theorem, we find
  $$
  {i\over 24\pi} \Gamma[q\Sigma_3] = \pm {1\over 32} \int_\Sigma d^2x \,
  \epsilon_{\mu\nu}\,{\rm STr}\,q\,\partial_\mu q\,\partial_\nu q \;.
  $$
Given the relation $S_5[q] = N W[q\Sigma_3]$ and the formula for
$W[M]$, comparison of this result with the nonlinear sigma model
action (\ref{NLsM}) yields
\begin{equation}
  \theta = \pm N\pi \;.
  \label{theta}
\end{equation}
The sign ambiguity arises from the multi-valuedness of $\Gamma[M]$.  It
has no consequence, since the physics of the nonlinear sigma model is
periodic in the topological angle $\theta$, as long as the position
space $\Sigma$ has no boundary.

For a complete description, what remains to be done is to augment the
effective action with a symmetry-breaking term for finite energy $E
\not= 0$.  By shifting $q \mapsto q - i(E/2\mu)\Sigma_3$ in
(\ref{eff_act}), and expanding to linear order in $E$, we obtain the
full effective action
\begin{equation}
  S_E[q] = S_{\rm eff}[q] - {i\mu NE \over 2g}
  \int d^2x \, {\rm STr} \, q\Sigma_3 \;.
  \label{fullaction}
\end{equation}
Finally, a standard calculation starting from the Lagrangian
(\ref{L0}) yields the following expression for the local density of
states:
\begin{equation}
  \nu(E;x) = {\mu N \over 2\pi g} {\rm Re} \int {\cal D}q\,{\rm Tr}
  \, q_{\rm FF}^{\vphantom{*}}(x)\sigma_3 \,{\rm e}^{-S_E[q]} \;,
  \label{ldos}
\end{equation}
where the functional integral has to be computed with a positive
imaginary part ${\rm Im}E > 0$, in the limit ${\rm Im}E \to 0$.

The above derivation of the couplings $f$ and $\theta$ was only of
formal validity, as the precise definition of the WZW model was
omitted and no justification of the bosonization rules was given.  In
the following two subsections, we are going to fill in the gaps.
Readers not interested in these details are urged to proceed directly
to Section \ref{sec:DOS}.

\subsection{WZW model of type $C|D$}
\label{sec:wzw}

In the present subsection we will construct the target space of the
WZW model, based on the general notion of Riemannian symmetric
superspace.  Although this construction is by no means new, it is not
as well known as it should be.  Many influential authors have based
their considerations on WZW models over the real supergroups ${\rm
  GL}_{\mathbb R} (n|n)$ or ${\rm OSp}_{\mathbb R}(2n|2n)$, or on the
unitary supergroups ${\rm U}(n|n)$.  While such practice may have
become standard, and is admittedly rather convenient and quite
satisfactory for a number of formal calculations in current algebra,
it does not take the WZW model seriously as a functional integral, and
we are not going to adopt it.  Bosonic ({\it i.e.}~non-super)
nonlinear sigma models, including the class of bosonic WZW models, are
defined as functional integrals of maps from a Riemann surface into a
target space with Riemannian structure.  If our purpose is to extend
this definition to the super world in a mathematically sound way, we
have to address the fact that the natural or invariant geometry on a
supergroup invariably is {\it non-Riemann}.

To appreciate the difficulty in some detail, consider ${\rm U}(n|n)$
for example.  The even part of the Lie superalgebra of ${\rm U}(n|n)$
is the direct sum of two copies of ${\rm u}(n)$, which is spanned by
the anti-Hermitian $n \times n$ matrices.  Thus, bosonic tangent
vectors at the unit element of ${\rm U}(n|n)$ are pairs $(A,B)$ with
$A^\dagger = -A$ and $B^\dagger = - B$.  The natural ${\rm U}(n|n)$
invariant quadratic form (or metric) evaluated on $(A,B)$ is the
supertrace ${\rm Tr}\, A^\dagger A - {\rm Tr} \, B^\dagger B$, which
has indefinite sign.  As a result, the action functional of the
principal chiral nonlinear sigma model with target supermanifold ${\rm
  U}(n|n)$ is bounded neither from below nor from above.  On field
configurations that fluctuate rapidly in space, the action becomes
arbitrarily large, and since the action can be either positive or
negative, the functional integral is unstable with respect to
fluctuations for any choice of sign of the coupling constant.
Therefore, unless some additional procedure (such as analytic
continuation from $A^\dagger A - B^\dagger B$ to $A^\dagger A +
B^\dagger B$) is specified, the theory with target ${\rm U}(n|n)$ does
not exist.

Such reasoning is not restricted to ${\rm U}(n|n)$ but holds for the
other cases as well.  Indeed, the Cartan-Killing form of any Lie
superalgebra is a supertrace, {\it i.e.}~a difference between two
traces.  For the supergroups listed above, this implies that any
rank-two (super)symmetric tensor $\kappa$ invariant under the left and
right group actions is necessarily non-Riemann, by which we mean that
the metric tensor obtained by restriction of $\kappa$ to the bosonic
support has indefinite signature.  For this universal reason,
supergroups are ruled out as a target spaces for nonlinear sigma
models, at least in the literal sense.

We understand that the Lagrangians of the ``supergroup'' WZW models
which exist in the literature, are used mainly for bookkeeping
purposes, or as a device to generate equations of motion or other
identities, such as the Knizhnik-Zamolodchikov equation, that do not
depend on working with a Riemannian geometry.  In the present paper,
we wish to put the WZW model (or some descendant theory) to a more
severe test: we intend to integrate carefully over the global zero
modes (treating the theory in zero-mode approximation), so as to
establish the connection with exact random-matrix limits that have
previously been obtained.  For that nonperturbative purpose, we must
construct a functional integral that exists as such.

While the basic difficulty with supergroup targets is easily
recognized, it is not so easy to circumvent.  For example, switching
from ${\rm U}(n|n)$ to its noncompact analog ${\rm GL}_{\mathbb C}
(n|n)/{\rm U}(n|n)$ does not improve the situation.  The latter
space is still non-Riemann, and the nonlinear sigma models over it
make sense only in combination with some procedure of analytic
continuation \cite{gll}. 

It turns out that the difficulty cannot be overcome by using only the
toolshed of standard supermanifold theory.  What is required is a
novel concept, namely that of {\it cs-manifold},\footnote{This
  terminology is due to Bernstein \cite{bernstein}.  The letter ``c''
  stands for complex, and ``s'' for super.} which transcends the
traditional categories of real-analytic and complex-analytic
supermanifolds.  In short, cs-manifolds are super bundles supported by
a real-analytic manifold, with a fibre which is a Grassmann algebra
over ${\mathbb C}$ carrying no operation of complex conjugation or
adjoint.  In plain physics language, the ``bosons are real while the
fermions are complex''.  It is not hard to see that this is just the
right kind of mathematical setting to use for the construction of
nonlinear sigma models with superspace target.  Indeed, what we want
is a target space with a {\it Riemannian metric} on its bosonic base
manifold, giving an action functional bounded from below.  A
Riemannian metric distinguishes the real numbers from the imaginary
numbers, in the sense that the tangent vectors with positive length
form a vector space over ${\mathbb R}$ (and not over $i{\mathbb R}$ or
${\mathbb C}$).  This is what is meant by saying that the ``bosons are
real''.  On the other hand, fermionic ``integration'' according to
Berezin is nothing but differentiation, and the definition of
Berezin's integral requires no notion of reality, so it is most
natural to ``leave the fermions complex'' \cite{ast}.  Our intention
as statistical physicists is to do integrals (computing disorder
averages of Green functions), and nothing but integrals.  Hence, we
may take the extreme point of view that complex conjugation, while
indispensable for the bosons, is a {\it forbidden} operation on
fermions.  This is the point of view advocated in \cite{rss}, and it
is the one we adopt here.  Note that this means that we bar
supergroups such as ${\rm U}(n|n)$, the definition of which requires
the use of an adjoint for the fermions.

Having prepared the stage with these remarks, we now turn to the
description of the target space of the WZW model.  Our starting point
is the complex supergroup $G \equiv {\rm OSp}_{\mathbb C}(2n|2n)$.
Its elements, which we denote by $M$, satisfy the equation $M^{-1} =
M^{\rm t}$, where the orthosymplectic transpose $M^{\rm t}$ was
defined by (\ref{transpose}) and $(M\phi)^{\rm t} = \phi^{\rm t}
M^{\rm t}$.  The group $G$ comes with a rank-two supersymmetric tensor
$\kappa = - {\rm STr} \, {\rm d}M^{-1} {\rm d}M = {\rm STr}\,(M^{-1}
{\rm d}M)^2$, which has the distinctive property of being invariant
under left and right translations $M \mapsto g_L^{\vphantom{-1}} M
g_R^{-1}$.  Let $G_0 = {\rm O}(2n, {\mathbb C}) \times {\rm Sp} (2n,
{\mathbb C})$ denote the ordinary (or bosonic) subgroup of $G$.  Since
$G_0$ is a group of matrices with complex entries, the geometry
induced on $G_0$ by restriction of $\kappa$ is non-Riemann.

What we need to do now is to specify a submanifold of $G_0$ on which
the geometry induced by $\kappa$ {\it is} Riemann.  This is done as
follows.  {}From ${\rm O}(2n,{\mathbb C})$ (the FF-sector) we select a
submanifold ${\cal M}_{\rm F}$ isomorphic to the compact orthogonal
group ${\rm O}(2n)$.  Near the group unit, ${\cal M}_{\rm F}$ is
parametrized by ${\rm e}^Y$ where $Y = -Y^\dagger = - \sigma_1 Y^{\rm
  T} \sigma_1$.  The group ${\rm O}(2n)$ acts on ${\cal M}_{\rm F}$
independently on the left and right by ${\rm e}^Y \mapsto O_L^
{\vphantom{-1}}\,{\rm e}^Y O_R^{-1}$, and this action is transitive,
which is to say that all elements of ${\cal M}_ {\rm F}$ are
translates of unity.  In the BB-sector we proceed differently, by
selecting the intersection, ${\cal M}_{\rm B}$, of ${\rm Sp}(2n,
{\mathbb C})$ with the set of positive Hermitian matrices.  The
elements of ${\cal M}_{\rm B}$ can be written as $g g^\dagger$ with $g
\in {\rm Sp}(2n,{\mathbb C})$.  Since forming the product $gg^\dagger$
divides out the maximal compact subgroup ${\rm Sp}(2n)$ of ${\rm
  Sp}(2n,{\mathbb C})$ on the right, the set ${\cal M}_{\rm B}$ is
isomorphic to ${\rm Sp}(2n, {\mathbb C}) / {\rm Sp}(2n)$.  Using the
exponential map, ${\cal M}_{\rm B}$ can be parametrized by ${\rm e}^X$
where $X = +X^\dagger = - \sigma_2 X^{\rm T} \sigma_2$.  Note that the
product of two positive Hermitian matrices is not positive Hermitian
in general, so ${\cal M}_{\rm B}$ is not a group.  Rather, ${\cal
  M}_{\rm B}$ is a noncompact symmetric space, and the complex group
${\rm Sp}(2n, {\mathbb C})$ acts on ${\cal M}_{\rm B}$ transitively by
${\rm e}^X \mapsto g \, {\rm e}^X g^\dagger$.

We now look at the product ${\cal M}_{\rm B} \times {\cal M}_{\rm F}$.
This may seem like a somewhat unnatural hybrid to consider, since
${\cal M}_{\rm F}$ is group whereas ${\cal M}_{\rm B}$ is not.
However, both ${\cal M}_{\rm F}$ and ${\cal M}_{\rm B}$ are Riemannian
symmetric spaces, the former of compact and the latter of noncompact
type, and the product ${\cal M}_{\rm B} \times {\cal M}_{\rm F}$ has
the desired property of being a Riemannian submanifold of $G_0$.  To
establish the Riemannian property, we first inspect the tangent space
at unity, ${\cal T}_1 \left( {\cal M}_{\rm B} \times {\cal M}_{\rm F}
\right)$.  Its elements are pairs $A \oplus B = {d \over ds} {\rm
  e}^{s(A \oplus B)} \Big|_{s=0}$, and evaluation of $\kappa = {\rm
  STr} \, (M^{-1} {\rm d}M)^2$ on these (at $M = 1$) yields
  $$
  {\rm STr} \left( A \oplus B \right)^2 = {\rm Tr} \, A^2 - {\rm Tr} \,
  B^2 = {\rm Tr} \, A^\dagger A + {\rm Tr} \, B^\dagger B \ge 0 \;.
  $$
{}From this we clearly see that $\kappa$ is positive definite on ${\cal
  T}_1 \left( {\cal M}_{\rm B} \times {\cal M}_{\rm F} \right)$.  This
property carries over to all of ${\cal M}_{\rm B} \times {\cal M}_{\rm
  F}$, by the invariance of $\kappa$ under the transitive action of
the group ${\rm Sp}(2n,{\mathbb C}) \times \left( {\rm O}(2n)_L \times
  {\rm O}(2n)_R \right)$.  Thus, $\kappa$ restricts to a Riemannian
structure on ${\cal M}_{\rm B} \times {\cal M}_{\rm F}$ as claimed.

As an immediate consequence, we have that the numerical part of the
function ${\rm STr}\,M = {\rm STr}\,M^{-1}$ on ${\cal M}_{\rm B}
\times {\cal M}_{\rm F}$ is locally expressed by
\begin{eqnarray}
  {\rm STr}\, M_0 &=& {\rm Tr}\,{\rm e}^A - {\rm Tr}\,{\rm e}^B
  = {\rm Tr}\,{\rm e}^{-A} - {\rm Tr}\,{\rm e}^{-B} \nonumber \\ &=&
  {\rm Tr}\,\cosh \sqrt{A^\dagger A} - {\rm Tr} \cos \sqrt{B^\dagger B} \;.
  \label{minimum}
\end{eqnarray}
This function has an absolute minimum at $M = 1$.

In summary, the object at hand is highly structured, consisting of the
complex supergroup $G = {\rm OSp}_{\mathbb C}(2n|2n)$ with metric
$\kappa = {\rm STr}\, (M^{-1}{\rm d}M)^2$, and a Riemannian
submanifold ${\cal M}_{\rm B} \times {\cal M}_{\rm F}$.  The triple
$(G,\kappa,{\cal M}_{\rm B}\times{\cal M}_{\rm F})$ is what is called
\cite{rss} a Riemannian symmetric superspace of type $C|D$.  The name
encodes the fact that the noncompact BB-sector is symplectic, or $C$,
while the compact FF-sector is orthogonal in even dimension, or $D$.

The virtue of Riemannian symmetric superspaces, as opposed to
supergroups, is that they make valid target spaces for nonlinear sigma
models, as follows.  Let the action functional be denoted by $S$ and
constructed in the usual way, {\it i.e.}~if $\kappa_{ij} (\varphi)
{\rm d}\varphi^i {\rm d}\varphi^j$ is the expression of the metric in
terms of target space supercoordinates $\varphi^i$, which are bosonic
for $i = 1,...,p$ and fermionic for $i = p+1,...,p+q$, we have $S =
\int d^2x \, \kappa_{ij} (\varphi) \partial_\mu \varphi^i \partial_\mu
\varphi^j$.  According to Berezin \cite{berezin} superintegration is a
two-step process.  First we do the Fermi integral, which is to say we
differentiate ${\rm e}^{-S}$ with respect to the fermionic fields.
(For this no definition of an adjoint or complex conjugation of the
fermions is needed.)  This ``integral'' always exists -- provided we
take care to control the infinite number of integration variables --
in the sense that the derivative of an analytic function always
exists.  The result of doing the Fermi integral is a functional, say
${\rm e}^{-S_0} D_0$, on the bosonic fields.  In the second step, we
carry out the integral over the bosonic fields, which take values in a
real target manifold (${\cal M}_{\rm B} \times {\cal M}_{\rm F}$ in
the present case).  This functional integral exists (modulo the
notorious need for regularization in the ultraviolet, regularization
of zero modes {\it etc.}) because we have constructed the target as a
Riemannian manifold, with a bosonic action $S_0$ bounded from below.

One might object that we are violating ``supersymmetry'' by using
complex conjugation on the bosons (to fix the Riemannian submanifold
${\cal M}_{\rm B} \times {\cal M}_{\rm F}$), while barring the use of
complex conjugation and any other adjoint on the fermions.  This is
not so.  The role of supersymmetry here is to equip the theory with a
BRST symmetry, thereby turning it into a kind of topological field
theory and ensuring normalization of the partition function to unity.
BRST symmetry does not require that we treat bosons and fermions in an
egalitarian manner with respect to complex conjugation.  The essence
of the argument can be captured by looking at a simple
zero-dimensional example.  For any analytic function $f : {\mathbb
  R}^+ \cup \{0\} \to {\mathbb C}$ with rapid decay at infinity and
$f(0) = 1$, consider the superintegral
  $$
  {1 \over 2\pi} \int_{{\mathbb R}^2} dx dy \,
  \partial_\xi \partial_\eta \, f(x^2 + y^2 + 2\xi\eta) = f(0) = 1 \;,
  $$
where $x$ and $y$ (the ``bosons'') are Cartesian coordinates of
${\mathbb R}^2$, and $\xi$ and $\eta$ (the ``fermions'') are Grassmann
variables.  The integral is always equal to unity, and this holds true
irrespective of whether any conjugation properties are imposed on
$\xi$ and $\eta$ or not.  (Thus we do not need to say that $\xi$ and
$\eta$ are ``real'', or $\eta$ is the ``complex conjugate'' of $\xi$,
or anything of that sort.)  The integral reduces to $f(0) = 1$ by
dimensional reduction \cite{ps}, as a consequence of invariance of the
integrand under BRST transformations
  $$
  \delta x = \varepsilon \xi \;, \quad
  \delta y = \tilde\varepsilon \eta \;, \quad
  \delta\xi = - \tilde\varepsilon y \;, \quad
  \delta\eta = \varepsilon x \;.
  $$
Since the generator of the BRST transformation vanishes only at the
origin $x = y = 0$, the integrand on ${\mathbb R}^2 \setminus (0,0)$
can be written as the BRST derivative of another function, causing
localization of the integral to the origin.  The same localization
principle applies in the functional setting.

One might ask again the basic question why we use the highly
structured concept of Riemannian symmetric superspace instead of more
conventional constructions (invoking an adjoint for both bosons and
fermions).  Isn't there a simpler way of doing it?  We have tried
hard, but apparently the answer is no, if we insist on global
supergroup symmetry and the stringent requirement of an invariant
Riemannian structure on the bosonic base of the target space.  There
is one thing, however, that we could do in the way of simplification
or economy of formulation, which would be to eliminate most of the
complex supergroup $G$ from the triple $(G,\kappa,{\cal M}_{\rm B}
\times {\cal M}_{\rm F})$, and retain just the Grassmann algebra
fibres over the points of the base ${\cal M}_{\rm B} \times {\cal
  M}_{\rm F}$.  The resulting object would be a cs-manifold in the
terminology of Bernstein.  We do not take this step here.

The WZW model whose existence we postulated earlier for the purpose of
expanding $S_5[q]$ in gradients, is the functional integral of maps
$M$ from position space into the Riemannian symmetric superspace of
type $C|D$,
  $$
  \Big( {\rm OSp}_{\mathbb C}(2n|2n), {\rm STr}(M^{-1}{\rm d}M)^2,
  {\cal M}_{\rm B} \times {\cal M}_{\rm F} \Big) \;,
  $$
with the action functional given in (\ref{wznw}).  (To make the model
completely well-defined, we add an infinitesimal mass term $\epsilon
\int d^2x \, {\rm STr}(M+M^{-1})$ to regularize the zero modes.)  We
propose to call it the $C|D$ WZW model at level $k = 1$.  The simpler
name ``${\rm OSp}(2n|2n)$'' WZW model would be inappropriate for two
reasons.  Firstly, it would belittle the above discussion emphasizing
the necessity to use a Riemannian symmetric superspace instead of a
supergroup target.  Secondly, such a name could cause some confusion
since there exists yet another ``${\rm OSp}(2n|2n)$'' WZW model, still
at level $k = 1$, which differs from the present one by the exchange
of the BB- and FF-sectors -- in other words, the bosons carry an
orthogonal structure while the fermions are symplectic.  (The latter
model, with $k = 2$, may have some relation to the fixed point
governing the spin quantum Hall transition \cite{smf,glr,bl}.)

There is one point that needs further attention.  Although the
existence and stability of the principal chiral nonlinear sigma model
with target $C|D$ should now be clear, the situation for the WZW model
is still precarious.  The reason is the presence of the multi-valued
term $\Gamma[M]$ in addition to the kinetic term induced by the metric
tensor $\kappa$.  While $i\Gamma[M]/24\pi$ is imaginary (and ambiguous
by integer multiples of $2\pi i$) in the FF-sector, it is {\it real}
(and single-valued) in the BB-sector.  Indeed, if $A = A^\dagger$, $B
= B^\dagger$ and $C = C^\dagger$ are elements of the tangent space of
${\cal M}_{\rm B}$ at unity, the 3-linear form $i {\rm Tr}(A[B,C])$,
which induces the BB-part of $i\Gamma[M]$, takes values in ${\mathbb
  R}$.  Moreover, this term reverses its sign under $(A,B,C) \mapsto
(-A,-B,-C)$, so there exist field directions along which ${\rm e}^{-i
  \Gamma[M]/24\pi}$ increases exponentially, thereby jeopardizing the
existence of the functional integral with action $W[M]$.  Fortunately,
one can prove the following bound:
\begin{equation}
  \left| {{\rm Re} \over 24\pi} \, i\Gamma[M] \right| < {{\rm Re}
    \over 16\pi} \int d^2x \, {\rm STr} (M^{-1}\partial_\mu M)^2 \;,
  \label{bound}
\end{equation}
which ensures the stability of the functional integral $\int{\rm
  e}^{-W[M]}$.  For the special case ${\cal M}_{\rm B}={\rm H}^3$
(three-hyperboloid), which is isomorphic to ${\cal M}_{\rm B}={\rm
  Sp}(2n,{\mathbb C})/{\rm Sp} (2n)$ for $n = 1$, this bound was
established in \cite{cftiqhe}.

As a final remark, let us mention that the bound (\ref{bound}) is {\it
optimal}.  In other words, if we multiply the kinetic term of the WZW
action $W[M]$ by a factor less than unity, and arbitrarily close to
unity, there exist field configurations violating the bound, and the
functional integral becomes unstable.  This means that the $C|D$ WZW
model resides {\it exactly on the border of stability}.  An important
consequence is that $W[M]$ tolerates the addition of a $J_L J_R$
current-current perturbation only with a definite sign of the
coupling.

\subsection{Nonabelian bosonization}
\label{sec:nonab}

With a good functional integral in hand, we can now go ahead and
establish a supersymmetric extension of nonabelian bosonization.  We
are going to show that the free Dirac theory plus $b$-$c$ ghost system
is equivalent to the $C|D$ WZW model at level $k = 1$.  To that end,
we consider the partition function
  $$
  {\cal Z}_{\rm WZW}[\bar A,A] = \int {\cal D}M \, \exp - W[M;\bar A,A] \;,
  $$
where $W[M;\bar A,A]$ is a gauged WZW action:
\begin{eqnarray*}
  W[M;\bar A,A] &=& W[M] - {1 \over 2\pi} \int d^2x \, {\rm STr} \Big(
    M^{-1}\bar\partial M A + \bar A M \partial M^{-1} \\
    &&\hspace{5cm} + \bar A M A M^{-1} - \bar A A \Big) \;,
\end{eqnarray*}
and $\bar A$ and $A$ are external sources taking values in ${\rm osp}
(2n|2n)$.  Our main tool for analysing the WZW model is the
Polyakov-Wiegmann relation \cite{pwrel}
\begin{equation}
  W[gh^{-1}] = W[g] + W[h^{-1}] - {1 \over 2\pi} \int d^2x \, {\rm STr}
  \, g^{-1}\bar\partial g \, h^{-1}\partial h \;.
  \label{pwrel}
\end{equation}
By using it, we easily derive the identity
  $$
  W[g_L^{\vphantom{-1}} M g_R^{-1}] = W[M;g_L^{-1} \bar\partial
  g_L^{\vphantom{-1}} , g_R^{-1} \partial g_R^{\vphantom{-1}}] +
  W[g_L^{\vphantom{-1}} g_R^{-1}]\;,
  $$
and parametrizing the external sources by
\begin{equation}
  \bar A = g_L^{-1} \bar\partial g_L^{\vphantom{-1}} \;,
  \quad A = g_R^{-1} \partial g_R^{\vphantom{-1}} \;,
  \label{sources}
\end{equation}
we can compute the partition function ${\cal Z}_{\rm WZW}[\bar A
,A]$ exactly:
  $$
  {\cal Z}_{\rm WZW}[\bar A,A] = \int {\cal D}M \, {\rm e}^{-
    W[g_L^{\vphantom{-1}} M g_R^{-1}] + W[g_L^{\vphantom{-1}} g_R^{-1}]}
  = \exp W[g_L^{\vphantom{-1}} g_R^{-1}] \;.
  $$
To arrive at the last expression, we changed variables from $M$ to
$M^\prime = g_L^{\vphantom{-1}} M g_R^{-1}$ in the functional
integral.  Such a substitution is certainly valid at the infinitesimal
level, {\it i.e.}~as long as the Taylor expansion of ${\cal Z}_{\rm
  WZW}[\bar A,A]$ with respect to the sources $\bar A, A$ is truncated
at finite order.

{}From the Polyakov-Wiegmann relation, one can also verify the
invariance of $W[M]$ under local ${\rm OSp}_{\mathbb C} (2n|2n)_L
\times {\rm OSp}_{\mathbb C}(2n|2n)_R$ transformations,
  $$
  M(z,\bar z)\mapsto g_L^{\vphantom{-1}}(z) M(z,\bar z)g_R(\bar z)^{-1}\;,
  $$
which is characteristic of any WZW model and takes the form of an
affine Lie symmetry in the quantum theory, which here is nonunitary.
(Again, the invariance holds without doubt at the infinitesimal level,
since we can always pass to the complexified tangent space with
impunity.)  By Noether's theorem, the invariance entails two sets of
conserved currents:
  $$
  J = M\partial M^{-1} \;, \qquad \bar J = M^{-1} \bar\partial M \;,
  $$
satisfying the equations of motion $\bar\partial J=\partial\bar J=0$.

Consider now the free Dirac theory plus $b$-$c$ ghost system,
with the partition function
\begin{eqnarray*}
  {\cal Z}_{\rm Dirac}[\bar A,A] &=& \int {\cal D}\phi \, {\cal D}
  \bar\phi\, \exp\, -\int d^2x \Big( \bar\phi^{\rm t} (\partial + A)
  \bar\phi + \phi^{\rm t} (\bar\partial + \bar A) \phi \Big) \\ &=&
  {\bf SDet}^{-1/2} \pmatrix{0 &\partial+A\cr \bar\partial+\bar A &0\cr}\;.
\end{eqnarray*}
This functional (super)determinant is ill-defined and needs to be
regularized in the infrared, say by adding to the Dirac Lagrangian
a small mass term,
  $$
  \epsilon \left( \bar\phi^{\rm t} \Sigma_3 \phi + \phi^{\rm t}
    \Sigma_3 \bar\phi \right) = 2\epsilon ( |b|^2 + |c|^2 + ... ) \;.
  $$
Using standard heat kernel techniques \cite{alvarez}, the regularized
superdeterminant can then be computed to be the exponential of another
WZW action:
  $$
  {\cal Z}_{\rm Dirac}[\bar A,A] =
  {\bf SDet}^{-1/2} \pmatrix{
    \epsilon\Sigma_3 &\partial+A\cr
    \bar\partial+\bar A &\epsilon\Sigma_3\cr}
  = \exp W[g_L^{\vphantom{-1}} \Sigma_3 g_R^{-1} \Sigma_3] \;,
  $$
where the sources $\bar A$ and $A$ were again assumed to be of the
form (\ref{sources}).  By comparing answers, we see that the partition
functions of the $C|D$ ${\rm WZW}_{k=1}$ model (derived from the
complex supergroup ${\rm OSp}(2n|2n)$) and the Dirac$+$ ghost system
(with $n$ species of particles) coincide:
  $$
  {\cal Z}_{\rm Dirac}[\bar A,A] =
  {\cal Z}_{\rm WZW}[\Sigma_3\bar A \Sigma_3,A] \;,
  $$
on gauge source fields of the form (\ref{sources}).  This equivalence
furnishes the basis for nonabelian bosonization.

As an immediate consequence we get bosonization rules for the
currents:
\begin{equation}
  \phi \phi^{\rm t} \sim {1 \over 2\pi} \partial M \, M^{-1} \;, \quad
  \Sigma_3 \bar\phi\bar\phi^{\rm t} \Sigma_3 \sim - {1 \over 2\pi}
  M^{-1}\bar\partial M \;,
  \label{boson_current}
\end{equation}
by comparing the $J\cdot A$ perturbations in the two theories.  (Using
these bosonization rules we can rewrite a single species of Dirac
fermions with random mass as a WZW model with marginally irrelevant
current-current interaction.  No additional insight seems to be gained 
from that alternative representation.)  However, in our computation of
the gradient expansion of $S_5[q]$, Eq.~(\ref{eff_act}), we employed
the bosonization rules
\begin{equation}
  \phi\bar \phi^{\rm t}\Sigma_3 \sim (2\pi\ell)^{-1} M \;, \quad {\rm
    and} \quad \Sigma_3\bar\phi \phi^{\rm t} \sim (2\pi\ell)^{-1} M^{-1} \;.
  \label{boson_dens}
\end{equation}
To justify these, further considerations are necessary.

The $C|D$ ${\rm WZW}_{k=1}$ model, as any WZW model, is solvable to a
high extent, owing to the availability of exact operator product
expansions between the conserved currents and the fundamental field
$M$ (and all other primary fields). To write these down, we introduce
the convenient notation $J^A = -{1\over 2} {\rm STr}\, A M \partial
M^{-1}$, where $A$ is any spatially constant element $A \in {\rm
osp}(2|2)$.  Then, making in the functional integral $\langle M(w,\bar
w) \rangle$ a change of variables $M \to {\rm e}^{- \epsilon X} M$,
where $X$ is holomorphic in a small neighborhood of the point $w$ and
smoothly goes to zero outside, we obtain the operator product
expansion \cite{kz}
\begin{equation}
  J^A(z) M(w,\bar w) = - {A \over z-w} M(w,\bar w) + ... \;.
  \label{opeJM}
\end{equation}
For the antiholomorphic current defined by $\bar J^A(\bar z)=-{1\over 2}
{\rm STr} A M^{-1}\bar\partial M$, an analogous calculation gives
\begin{equation}
  \bar J^A(\bar z) M(w,\bar w) = M(w,\bar w){A\over\bar z-\bar w}+...\;.
  \label{opebJM}
\end{equation}
The OPE of the conserved currents among themselves can be obtained in
a similar manner.  For the holomorphic current we have
\begin{equation}
  J^A(z) J^{B}(w) = - {{1\over 2}{\rm STr}\, AB \over (z-w)^2}
  + {J^{[A,B]}(w) \over z-w} + ... \;,
  \label{kacmoody}
\end{equation}
and the same formula (with $z \to \bar z$ {\it etc.}) holds for the
antiholomorphic one.

On general grounds, the holomorphic component $T$ of the stress-energy
tensor has the canonical Sugawara form $T(z) \propto \kappa_{ij} :
J^i(z) J^j(z) :$ , where $J^i = - {1\over 2}{\rm STr}\, e^i M \partial
M^{-1}$, and $\{e^i\}$ is a basis of the Lie superalgebra ${\rm
  osp}(2n,2n)$.  The metric tensor is expressed by $\kappa_{ij}=-{\rm
  STr}\,e_i e_j$, indices being lowered via $\delta_j^i = {\rm STr}\,
e^i e_j$.  Classical considerations (by the construction of $T$ from
the Lagrangian via Legendre transform) would suggest a constant of
proportionality $1/k$, where $k = 1$ in the present case.  However,
the constant is renormalized by quantum fluctuations.  Its correct
value is deduced from the requirement
  $$
  T(z) J^A(w) = {J^A(w) \over (z-w)^2} + ... \;,
  $$
expressing the fact that $J$ is a holomorphic conserved current
and therefore must have conformal dimensions $(\Delta,\bar\Delta) =
(1,0)$.  Expanding the product $T(z) J^A(w)$ with the help of
(\ref{kacmoody}) and the associativity of the operator product
algebra, one finds
\begin{equation}
  T(z) = {\kappa_{ij} \over 1 + h_*} : J^i(z) J^j(z) : \;,
  \label{sugawara}
\end{equation}
where $h_*$, called the dual Coxeter number, is the quadratic
Casimir invariant evaluated in the adjoint representation:
  $$
  \kappa_{ij} [ e^i , [ e^j , A ] ] = h_* A \;.
  $$
The superbracket $[\cdot,\cdot]$ here means the commutator or the
anticommutator, as is appropriate.  A standard calculation using the
root system of ${\rm osp}(2n,2n)$ yields
  $$
  h_* = - 2 \;,
  $$
independent of $n$.  Note that $h_*$ is negative!  We will see shortly
that this has to be so, in order for the dimensions of operators such
as the fundamental field $M$ to come out being positive.  (We can also
check the sign of $h_*$ by interpreting the $C|D$ WZW model with say,
$n = 1$, as the replica limit $r \to 0$ of the familiar ${\rm O}(r)$
WZW model.  As is well-known, the Lie algebra of ${\rm O}(r)$ has dual
Coxeter number $h_* = r - 2$, which indeed becomes $-2$ at $r = 0$.)
The Fourier modes of $T(z)$,
  $$
  L_m = (2\pi i)^{-1} \oint T(z)\, z^{m+1} dz \;,
  $$
are the generators of a Virasoro algebra, with the central charge $c$
being zero by supersymmetry.  Similar statements hold for the
antiholomorphic component $\bar T(\bar z)$ of the stress-energy
tensor.  The presence of these Virasoro algebras implies the conformal
invariance of the WZW model.  From equations (\ref{opeJM})--(\ref{sugawara})
one sees that the leading singularities in the operator product expansion
of the stress-energy tensor with the fundamental field $M$ are
\begin{eqnarray*}
  T(z) M(w,\bar w) &=& {\Delta_M \over (z-w)^2} M(w,\bar w) + ... \;, \\
  \bar T(\bar z) M(w,\bar w) &=& {\bar\Delta_M \over (\bar z-\bar w)^2}
  M(w,\bar w) + ... \;.
\end{eqnarray*}
where
  $$
  \Delta_M = \bar\Delta_M = {C_M \over 1 + h_*} \;,
  $$
and $C_M = \kappa_{ij} e^i e^j$ is the quadratic Casimir in the
fundamental representation of ${\rm osp}(2n|2n)$.  The numbers
($\Delta_M , \bar\Delta_M$) are the conformal dimensions of the field
$M$.  We can calculate $C_M$ by making an explicit choice of basis
$\{e^i\}$.  In this way we find $C_M = -1/2$, which yields a total
dimension of one:
  $$
  {\rm dim}(M) = \Delta_M + \bar\Delta_M =
  2 \times {C_M \over 1 + h_*} = 2 \times {-1/2 \over 1 - 2} = 1 \;.
  $$
This means that the two-point function $\langle M M \rangle$ must
vary as the inverse square of the distance.  Global $G_L \times
G_R$ invariance then constrains the OPE of $M$ with itself to be
of the form
  $$
  {\rm STr}\, A M(z,\bar z) \times {\rm STr}\, B M^{-1}(w,\bar w)
  = - {\ell^2 \over (2\pi |z - w|)^2} \, {\rm STr}\, AB + ... \;,
  $$
where $\ell$ is a length scale depending on the choice of UV
regularization scheme.  The supermatrices $A$ and $B$ are arbitrary
elements of ${\rm osp}(2n|2n)$.

We now return to the Dirac$+$ghost system, and rewrite the OPE given
in (\ref{opedirac}) in the invariant form
  $$
  {\rm STr}\, A \phi(z) \bar\phi^{\rm t}(\bar z) \Sigma_3 \times
  {\rm STr}\, B \Sigma_3 \bar\phi(\bar w) \phi^{\rm t}(w) =
  - {{\rm STr}\, AB \over (2\pi|z-w|)^2} + ... \;.
  $$
(The presence of the factors $\Sigma_3$ leads to the overall minus
sign on the right-hand side.)  By comparing the two OPEs we see that
they agree to leading order if we identify $M/\ell$ with $\phi\bar
\phi^{\rm t} \Sigma_3$, and $M^{-1}/\ell$ with $\Sigma_3\bar\phi
\phi^{\rm t}$.  Moreover, the OPEs between these fields and the
respective currents agree to leading order in the two theories.  By
computing the three- and four-point functions one finds \cite{kz} that
the agreement persists to higher order.  This then justifies the
bosonization rules (\ref{boson_dens}).  Their status is the same as in
the usual case \cite{vdp}.

In summary, we have established the $C|D$ ${\rm WZW}_{k=1}$ model to
be equivalent to the Dirac plus $b$-$c$ ghost system {\it at the
  current algebra (or infinitesimal) level}.  What about aspects that
transcend the infinitesimal level?  This is not an empty question, as
the Riemannian symmetric superspace of type $C|D$ with bosonic
submanifold ${\cal M}_{\rm B} \times {\cal M}_{\rm F} = ({\rm
  Sp}(2n,{\mathbb C})/{\rm Sp}(2n)) \times {\rm O}(2n)$ has {\it two}
connected components, owing to the topological equivalence ${\rm
  O}(2n) \simeq {\mathbb Z}_2 \times {\rm SO}(2n)$.  Since the
discreteness of ${\mathbb Z}_2$ allows the WZW field to break up into
domains separated by domain walls, we are led to ask: exactly how is
the functional integral to be defined?  Do we have to sum over all
possible numbers and positions of the domain walls, or do we not?

We believe that the answer to the question is yes, and the precise
nonperturbative definition of the functional integral does have to
involve ${\mathbb Z}_2$ as a {\it local} degree of freedom, for the
following reason.  ${\mathbb Z}_2$ already exists in the classical
setting, and is not a peculiarity of the supersymmetric formalism.
Indeed, according to Witten \cite{witten}, the free 2d massless
$n$-component Majorana fermion theory is equivalent to the level-one
WZW model not over ${\rm SO}(n)$ but over the disconnected group ${\rm
  O}(n)$.  The argument given in \cite{witten} proceeds by the
comparison of current algebras, and does not address the issue of
domains and domain walls in the ${\mathbb Z}_2$ degree of freedom of
${\rm O}(n)$.  However, by specializing to the case $n = 1$, we see
that ${\mathbb Z}_2$ does need to be summed over locally, for a single
Majorana fermion is not an empty theory.  Rather, its partition
function bosonizes to that of the 2d Ising model, which is a theory of
local ${\mathbb Z}_2$ spin degrees of freedom.  In the continuum limit
near criticality, the latter partition function is computed by summing
over all possible numbers of domain walls and their positions.  By
counting the number of degrees of freedom we see that the situation is
the same for $n > 1$: the fermionic Hilbert space contains states that
cannot be produced by acting with currents on the vacuum.  To achieve
equality of the partition functions, one needs ${\mathbb Z}_2$ in the
WZW model, and ${\mathbb Z}_2$ must be summed over as a local spin
degree of freedom.  We particularly emphasize this point, although we
will not pursue it seriously in the present paper.

Here ends our extensive excursion into nonabelian bosonization and the
$C|D$ WZW model.  This material was included to make the present paper
self-contained.

\section{Density of states}
\label{sec:DOS}

In Section \ref{sec:Qfield} we saw that, when fluctuations are
neglected, the composite field $Q$ assumes a nonzero saddle-point
value $\mu$.  Since $Q$ enters the theory by coupling to the Dirac
bilinear $\phi_l^{\vphantom{\rm t}} \bar\phi_l^{\rm t} + \bar\phi_l^
{\vphantom{\rm t}}\phi_l^{\rm t}$, the latter acquires a nonvanishing
expectation value, too, and the density of states at zero energy
becomes finite.  This happens in spite of the fact that the Dirac
theory is devoid of any scale, and is an instance of dynamical mass
generation.  Of course, a fixed value of the ``order parameter'' $Q$
breaks the ${\rm OSp}(2n|2n)$ symmetry and leads to the existence of
Goldstone modes, the low-energy effective field theory for which is
the nonlinear sigma model (\ref{NLsM}).

What is the effect of order parameter fluctuations?  The
Mermin-Wagner-Coleman theorem states that continuous symmetries cannot
be spontaneous\-ly broken in two dimensions.  This statement applies
to {\it compact} symmetries, or nonlinear sigma models with compact
target spaces.  These flow under renormalization to strong coupling,
so that the field fluctuations grow large, and the potentially broken
symmetry is restored in the infrared.  However, for nonlinear sigma
models with supersymmetry, or zero replica number, another scenario is
possible, as has long been known from the example of time-reversal
invariant disordered 2d electron systems with spin-orbit scattering
(class $A{\rm II}$).  In that case, the beta function at weak coupling
has the ``wrong'' sign, which physically corresponds to weak {\it
  anti}localization, and the nonlinear sigma model undergoes
logarithmic flow to a Gaussian fixed point describing a perfect metal.
As was already mentioned in \cite{bcsz2}, the same happens in the
present case.  Let us review how to arrive at this result.

\subsection{Renormalization group}
\label{sec:RG}

Friedan \cite{friedan} has shown in great generality that the one-loop
beta function of a nonlinear model with target space metric $\kappa$
is determined by the Ricci curvature $R$:
  $$
  {d \kappa \over d\ln\ell} = - {R \over 2\pi} + ... \;.
  $$
The metric of the present target space is $\kappa = - (8\pi f)^{-1}
{\rm STr}\,({\rm d}q)^2$, and its Ricci curvature can be described as
follows.  (Note that the topological term of the nonlinear sigma model
does not renormalize at weak coupling, and can safely be ignored.  We
shall also neglect the nonperturbative ${\mathbb Z}_2$ degree of
freedom of the target space.)  We parametrize $q$ by the exponential
map, $q = {\rm e}^X \Sigma_3 {\rm e}^{-X}$ with $\Sigma_3 X \Sigma_3 =
- X$, and denote the commutator (or adjoint) action by $[X,\bullet]
\equiv {\rm ad}(X)$.  Then the Ricci curvature is the second-rank
invariant tensor determined by the quadratic form \cite{helgason}
  $$
  R_0(X,X^\prime) = - {\rm STr}\, {\rm ad}(X) {\rm ad}(X^\prime)
  $$
on the tangent space at $X = 0$, or $q = \Sigma_3$.  For any
irreducible symmetric space, this tensor will always be a constant
multiple of the metric tensor, as follows on general differential
geometric grounds and is necessary for the nonlinear sigma model to be
renormalizable.  Moreover, the constant of proportionality is
independent of $n$ by supersymmetry, and we may calculate it by
looking at the simplest case $n = 1$.  To do so, one evaluates both
tensors on some element of the tangent space, say $H = h \, E_{\rm BB}
\otimes \sigma_1$.  The eigenvalues of the adjoint action are called
{\it roots}.  In the present case with $n = 1$, there exist two
nonvanishing roots, a bosonic root $2h$ with multiplicity one, and a
fermionic root $h$ with multiplicity two.  Hence,
  $$
  - R_0(H,H) = (2h)^2 - 2 h^2 = 2 h^2 = {\rm STr}\, H^2 \;,
  $$
which gives $R = {\rm STr}\,({\rm d}q)^2 / 4$.  In terms of the
coupling $f$, the one-loop RG equation then reads
\begin{equation}
  {df \over d\ln\ell} = - f^2 \;.
  \label{RGflow}
\end{equation}
By integrating this equation from the cutoff scale $\ell_0$ to $\ell$,
with initial condition
  $$
  f_0 \equiv f(\ell_0) = N^{-1} (1 - {\rm e}^{-\pi/g})^{-1} \;,
  $$
we find that the coupling $f$ flows to zero as an inverse logarithm:
\begin{equation}
  f(\ell) = { f_0 \over 1 + f_0 \ln(\ell/\ell_0) } \;.
\label{runcoupl}
\end{equation}
This results in the system being a perfect ``metal'', with a
resistivity $(\sim f)$ that goes to zero in the thermodynamic limit.
By the same token, the broken ${\rm OSp}(2n |2n)$ symmetry will {\it
not} be restored, but remains truly broken, and the density of states
at $E = 0$ is sure to be nonzero.

As was mentioned earlier, spontaneous breaking of continous ${\rm
  OSp}(2n|2n)$ symmetry also occurs for 2d metals with spin-orbit
scattering (class $A{\rm II}$).  There is, however, one important
difference we wish to mention.  For class $A{\rm II}$ a metallic phase
does {\it not} exist in disordered wires (the quasi-1d limit), where
weak antilocalization at short scales always crosses over to strong
localization at large scales.  In contrast, for class $D$ anomalous
``metallic'' behavior already occurs in quasi-1d systems!  By solving
the $C{\rm I}|D{\rm III}$ nonlinear sigma model using its quantum
Hamiltonian, one finds that the (thermal) conductance decays
algebraically for wires of arbitrary length.  The anomalous behavior
can also be seen on a more elementary level from the maximum entropy
transfer matrix ensemble, evolving according to the so-called DMPK
equation, for class $D$.\footnote{We thank John Chalker for pointing
  this out. See also \cite{bfgm}.}

\subsection{RG for the density of states}
\label{sec:rgDOS}

We now embark on a quantitative calculation, based on (\ref{ldos}), of
the local density of states in the metallic limit.  The idea is to
follow the flow of the renormalization group starting from the
short-distance cutoff $\ell_0$ all the way up to the system size $L$,
and then evaluate the functional integral (\ref{ldos}) in
zero-dimensional approximation.  (The local ${\mathbb Z}_2$ degree of
freedom of the nonlinear sigma model is expected to play no essential
role in that process, and will neglected but for its global part.)
The initial cutoff is taken to be the length associated with the
dynamically generated mass, $\ell_0 \sim \mu^{-1}$, which sets the
scale over which the ``ballistic'' free fermion theory crosses over to
the nonlinear sigma model describing diffusion.  (From the nonabelian
bosonization argument given at the beginning of Section
\ref{sec:gradient}, we saw that the reduction to the nonlinear sigma
model is valid on length scales larger than $\mu^{-1}$.)

To begin, we add a source term $\lambda\,{\rm Tr}\,q_{\rm FF}^
{\vphantom{\dagger}}(0) \sigma_3$ to the action functional and
differentiate with respect to $\lambda$ at $\lambda = 0$.  From
\cite{friedan}, the parameter $\lambda$ obeys the renormalization
group equation
  $$
  {d\lambda \over d\ln\ell} = - f C_q \lambda \;,
  $$
where $C_q$ is the quadratic Casimir invariant of ${\rm osp}(2n|2n)$,
normalized according to the metric $-{\rm STr}\,({\rm d}q)^2/8$ and
evaluated on the representation the field $q$ transforms under.
This number turns out to be negative:
  $$
  C_q = - 1 \;,
  $$
so that $\lambda$ grows with increasing scale.  In the thermodynamic
limit, this will lead to a logarithmically divergent density of states,
as follows.

Let the system be finite and of size $L$.  By integrating the flow
equation for $\lambda$ from the microscopic cutoff $\ell_0$ up to $L$,
  $$
  \lambda(L) = \lambda(\ell_0) \exp \int_{\ell_0}^L f(\ell^\prime)
  d \ln \ell^\prime \;,
  $$
and inserting the scale dependence (\ref{runcoupl}) of the running
coupling $f$, we obtain
  $$
  \lambda(L) / \lambda(\ell_0) = 1 + f_0 \ln (L / \ell_0) \;.
  $$
To keep track of the renormalization of the operator $\int d^2x \,
{\rm STr} \, q \Sigma_3$ in the action functional $S_E$, it is
convenient to introduce a running energy parameter $\epsilon(\ell)$,
with initial value $\epsilon_0 = \mu NE / 2g$.  This variable, as
compared to $\lambda$, carries two extra dimensions from $d^2x$, and
therefore evolves according to the equation
  $$
  {d \epsilon \over d\ln\ell} = (2 + f)\,\epsilon \;,
  $$
which integrates to
  $$
  \epsilon(L) / \epsilon_0 = (L/\ell_0)^2
  \left(1 + f_0 \ln (L/\ell_0) \right) \;.
  $$
Given (\ref{ldos}), all this leads to a scaling relation:
\begin{equation}
  \nu(E,f_0)_L = \left( 1 + f_0 \ln (L / \ell_0) \right)
  \nu\left(E \epsilon(L)/\epsilon_0,f(L)\right)_{\ell_0} \;.
  \label{scalrel}
\end{equation}
Here $\nu(E,f_0)_L$ is the local density of states of a system of size
$L$, energy $E$, and nonlinear sigma model coupling $f_0$, with the
short-distance cutoff being understood to be $\ell_0$.  For $L \gg
\ell_0$ and $E = 0$, we immediately obtain a logarithmic law,
  $$
  \nu(0,f_0)_L \sim f_0 \ln(L/\ell_0) \;,
  $$
which is a result given before by Senthil and Fisher \cite{sfD}.

Actually, we can work out a more precise answer. The function
$\nu(.)_{\ell_0}$ on the right-hand side of (\ref{scalrel}) is meant
to be evaluated for a system whose size $L$ equals the cutoff
$\ell_0$.  Under such circumstances, the functional integral can be
calculated by retaining only the spatially homogeneous mode $q(x) =
q_0$ (zero-mode approximation):
\begin{equation}
  \nu(E)_{\ell_0} = {\mu N \over 2\pi g} \, {\rm Re} \int Dq_0 \,
  {\rm Tr}(q_{\rm FF}^{\vphantom{\dagger}} \sigma_3) \exp \left(
    {i\mu N E \ell_0^2 \over 2 g} \,{\rm STr} \, q_0 \Sigma_3 \right) \;,
\label{zeromode}
\end{equation}
which no longer depends on the coupling $f$.  The computation of this
integral is the subject of the next section.  Using the answer given
below in (\ref{rmtdos}), we have
\begin{equation}
  \nu(E,f_0)_L = \bar\nu_L + {\sin(2\pi\bar\nu_L E L^2) \over 2\pi E L^2} \;,
  \label{dosTL}
\end{equation}
where
  $$
  \bar\nu_L = {\mu N \over \pi g}
  \left( 1 + f_0 \ln(L/\ell_0) \right) \;.
  $$
Since we have neglected the influence of finite $E$ on the RG flow,
the validity of this result is restricted to small energies $E \ll
E_{\rm Th}$.  By standard reasoning, the relevant energy scale is the
Thouless energy $E_{\rm Th} = D / L^2$, where $D = g / (2\pi\mu N
f_0)$ has the meaning of a diffusion constant.  In the opposite regime
$E \gg E_{\rm Th}$, direct use of one-loop perturbation theory yields
\begin{eqnarray}
  \nu(E) &=& \bar\nu_{\ell_0} + {1 \over \pi} {\rm Re} \int
  {d^2k \over (2\pi)^2} \, {1 \over Dk^2 - 2iE} \nonumber \\
  &=& \bar\nu_{\ell_0} + {1 \over 8\pi^2 D} \ln \left(
    1 + \left( {D \over 2E \ell_0^2} \right)^2 \right) \;,
  \label{dos1L}
\end{eqnarray}
where the momentum integral was cut off by $|k| < 1/\ell_0$.  We
observe that the energy-dependent part of the density of states
behaves as $\ln(1/E)$ on intermediate scales, and as $E^{-2}$ in the
asymptotic regime of large $E$.

\subsection{Random matrix limit}
\label{sec:rmt}

For the case $n = 1$, we are now going to compute the superintegral
  $$
  I(\epsilon) = {\textstyle{1 \over 2}}\int Dq \,
  {\rm Tr} (q_{\rm FF}^{\vphantom{\dagger}} \Sigma_3) \exp
  {\textstyle{1 \over 4}} i\epsilon\,{\rm STr}\,q\Sigma_3 \;,
  $$
where $Dq$ is the invariant Berezin measure on the Riemannian
symmetric superspace ${\bf X}_1$ (Section \ref{sec:saddles})
normalized by $\int Dq\,\exp\,{1\over 4}i\epsilon\,{\rm STr} \,
q\Sigma_3 = 1$.  This integral appeared in (\ref{zeromode}), and can
be viewed as the density of states in the limit of small system size
(also referred to as the ergodic regime, or the universal random
matrix limit).  Although $I(\epsilon)$ can be shown to be
semiclassically exact, which is to say that $I(\epsilon)$ is
calculated exactly in saddle-point approximation (by a supersymmetric
generalization of the Duistermaat-Heckman theorem \cite{dh}), we shall
not make any use of this deep fact and proceed in a more pedestrian
way.

The first step is to write down an explicit parametrization for the
supermatrix $q$.  To prepare that step, it is convenient to change the
arrangement of the multiplets $\phi,\phi^{\rm t}$ to
  $$
  \phi = \pmatrix{\psi_-\cr c\cr  \psi_+\cr  b \cr} \;, \quad
  \phi^{\rm t} = (\psi_+ , b , \psi_- , - c  ) \;.
  $$
Next recall the definition of the orthosymplectic transpose of a
supermatrix $X$ by $(X\phi)^{\rm t} = \phi^{\rm t} X^{\rm t}$.
Elements of the Lie algebra ${\rm osp}(2|2)$ are solutions of the
equation $X = - X^{\rm t}$, and are easily verified to be of the form
\begin{equation}
  X = \pmatrix{a &\alpha &0 &-\beta\cr \delta &d &\beta &b\cr
    0 &\gamma &-a &-\delta\cr \gamma &c &\alpha &-d \cr} \;,
  \label{osp}
\end{equation}
where Roman letters stand for commuting numbers, while Greek letters
denote Grassmann variables.

Now, as was emphasized in Section \ref{sec:saddles}, the superspace
${\bf X}_1$ consists of two disjoint pieces.  The first contains the
diagonal matrix $\Sigma_3 \in {\rm osp}(2|2)$, which rearranges to
  $$
  \Sigma_3 = {\rm diag}(1,1,-1,-1) \;.
  $$
On this piece of ${\bf X}_1$, the supermatrix $q$ will be parametrized
as
  $$
  q^{(1)} = \Sigma_3 + 2
  \pmatrix{Z\tilde Z(1 - Z\tilde Z)^{-1}&-Z(1-\tilde ZZ)^{-1}\cr
    \tilde Z(1-Z\tilde Z)^{-1} &-\tilde ZZ(1-\tilde ZZ)^{-1} \cr} \;.
  $$
This can be seen to obey the nonlinear constraint $q^2 = 1$.  From
(\ref{osp}) and the fact that $q = T \Sigma_3 T^{-1}$ lies in ${\rm
  osp}(2|2)$, the $2 \times 2$ supermatrices $Z$ and $\tilde Z$ have
to be of the form
  $$
  Z = \pmatrix{0 &-\beta \cr \beta & b} \;, \qquad
  \tilde Z = \pmatrix{0 & \gamma \cr \gamma & c} \;.
  $$
Following Section \ref{sec:saddles}, the bosonic base of ${\bf X}_1$
is fixed by the conditions
  $$
  b = c^* \;, \quad |b|^2 < 1 \;.
  $$

The second piece of ${\bf X}_1$ contains the diagonal matrix ${\rm
  diag}(-1,1,1,-1)$, which can be represented as
  $$
  {\rm diag}(-1,1,1,-1) = O \Sigma_3 O^{-1}
  $$
with
  $$
  O = E_{\rm FF} \otimes \sigma_1 + E_{\rm BB} \otimes 1_2 =
  \pmatrix{0 &0 &1 &0\cr 0 &1 &0 &0\cr 1 &0 &0 &0\cr 0 &0 &0 &1\cr}
  \in {\rm OSp}(2|2) \;.
  $$
Since $O$ lies in the orthosymplectic Lie supergroup, the $q$
matrix for the second piece of ${\bf X}_1$ can be obtained from the
first one by conjugation with $O$:
  $$
  q^{(2)} = T^\prime O \Sigma_3 O^{-1} {T^\prime}^{-1} =
  O T \Sigma_3 T^{-1} O^{-1} = O q^{(1)} O^{-1} \;.
  $$

The next step is express the integration measure $Dq$ in terms of $Z$
and $\tilde Z$.  One way of doing it is to use the coordinate
expression of the metric ${\rm STr}\,({\rm d}q)^2$.  A straightforward
calculation yields
  $$
  -{\rm STr}\,({\rm d}q)^2 / 8 = {\rm STr} \,(1-\tilde Z Z)^{-1}
  {\rm d}\tilde Z (1-Z \tilde Z)^{-1} {\rm d}Z \;.
  $$
This now gives rise to a Berezin measure in the standard way, {\it cf.}
Appendix $F$ of Ref.~\cite{bcsz1}, where a similar calculation was
described in full detail for symmetry class $C$.  One finds
  $$
  Dq = D(Z, \tilde Z)\, {\rm SDet}\, (1-Z \tilde Z)^{-1} \;,
  $$
where $D(Z,\tilde Z)$ is a flat Berezin measure.  On setting $b = r
{\rm e}^{i \varphi}$, we arrive at the coordinate expression
  $$
  Dq = (2\pi)^{-1} r {\rm d}r \wedge {\rm d}\varphi
  \,\partial_\beta \partial_\gamma\circ(1-r^2-2\beta\gamma)^{-1} \;,
  $$
which has been normalized so as to satisfy $\int Dq \, \exp \,
{1\over 4}i\epsilon \, {\rm STr} \, q\Sigma_3 = 1$.

We are finally ready to compute $I(\epsilon)$.  Using
\begin{eqnarray*}
  {\textstyle{1 \over 4}}{\rm STr}\, q^{(1)} \Sigma_3 &=&
  {r^2 \over 1 - r^2} + {2\beta\gamma \over (1 - r^2)^2} \;, \\
  {\textstyle{1 \over 4}}{\rm STr}\, q^{(2)} \Sigma_3 &=&
  {1 \over 1 - r^2} + {2 r^2 \beta\gamma \over (1 - r^2)^2} \;, \\
  {\textstyle{1 \over 2}}{\rm Tr}\, q_{\rm FF}^{(1)} \sigma_3
  &=& 1 - {2\beta\gamma \over 1 - r^2} =
  - {\textstyle{1 \over 2}}{\rm Tr}\, q_{\rm FF}^{(2)} \sigma_3 \;,
\end{eqnarray*}
and making the substitution $t = (1-r^2)^{-1}$, we obtain the following
expression for the integral:
\begin{eqnarray*}
  I(\epsilon) &=& \int_1^\infty {dt \over 2t^2}\,\partial_\beta
  \partial_\gamma\,(t + 2t^2\beta\gamma)(1 - 2t\beta\gamma)\times\cr
  \noalign{\vskip8pt}
  &\times&\biggl(\exp{\left(i\epsilon\left(t-1+2t^2\beta\gamma\right)\right)}
  - \exp{\left(i\epsilon\left(t+2t(t-1)\beta\gamma\right)\right)}\biggr)\;.
\end{eqnarray*}
where we have combined the contributions from the two pieces of ${\bf
  X}_1$.  The integral is now easily calculated to be
\begin{equation}
  I(\epsilon) = 1 - {1 \over i\epsilon} +
  {{\rm e}^{i\epsilon} \over i\epsilon} \;.
  \label{rmtdos}
\end{equation}
On taking the real part, we get
  $$
  {\rm Re}\,I(\epsilon) = 1 + {\sin\epsilon \over \epsilon} \;,
  $$
in agreement with the known result \cite{az} for the density of states
(in scaled units) of the Gaussian random matrix ensemble for class $D$.

Let us remark that exactly the same result for $I(\epsilon)$ would
have been obtained by treating the superintegral in saddle-point
approximation, in the spirit of the Duister\-maat-Heckman theorem.
The constant part comes from a saddle point on the trivial component
(with respect to ${\mathbb Z}_2$) of the target space, and the
oscillatory part ${\rm e}^{i\epsilon}/i\epsilon$ from a saddle point
on the nontrivial component.

\section{Disordered $d$-wave superconductor}
\label{sec:sc}

An important physical realization of symmetry class $D$ is by the
low-energy quasiparticles of dirty superconductors with broken
spin-rotation and time-reversal symmetry.  Although there is a number
of interesting cases to consider \cite{sfD,rg}, we will here focus on
a specific model of a disordered $d$-wave superconductor.  The
treatment in this section follows Ref.~\cite{asz} to some extent.  We
shall work in two-dimensional space with Cartesian coordinates $x, y$
and wave vector $k = (k_x,k_y)$.

The effective quasiparticle Hamiltonian of the pure system is
\begin{equation}
  H_0 = \sum_{k,\sigma} (\varepsilon_k - \mu) \, c_{k\sigma}^\dagger
  c_{k\sigma}^{\vphantom{\dagger}} + \sum_k \Delta_k \big(
    c_{k\uparrow}^\dagger c_{-k\downarrow}^\dagger + {\rm h.c.} \big) \;,
    \label{BdG}
\end{equation}
where the single-particle energies $\varepsilon_k$, shifted by the
chemical potential $\mu$, and the gap function $\Delta_k$ are taken
to be
\begin{eqnarray*}
  \varepsilon_k &=& - t \left( \cos k_x a + \cos k_y a \right) \;, \\
  \Delta_k &=& - \Delta_0 \left( \cos k_x a - \cos k_y a \right) \;,
\end{eqnarray*}
and $a$ is the lattice constant of the tight-binding model that
results on transforming to position space.  The excitations of this
Hamiltonian are gapped almost everywhere in the wave vector plane, the
exceptional places being the four points given by $|k_x| = |k_y| =
a^{-1} \arccos (-\mu / 2t)$, where $\varepsilon_k - \mu = \Delta_k =
0$.  Since the low-temperature behavior of the system on large
distance scales is expected to be dominated by the quasiparticles with
low energy, it is natural to linearize $H_0$ around the nodal points
$k_{x,y} = \pm a^{-1} \arccos(-\mu/2t)$, denoted in a self-explanatory
notation by $(\pm \pm)$.  By following a standard procedure
\cite{fradkin,ntw,sl}, we obtain four Majorana Hamiltonians, one for
each node. The one for $(++)$ reads
\begin{eqnarray*}
  H_0^{(++)} &=& {\textstyle{1\over 2}}iv_1\int d^2x\sum_a\left(f_a^\dagger
    \partial_1^{\vphantom{\dagger}} f_a^\dagger + f_a \partial_1 f_a \right)\\
  &+& {\textstyle{1\over 2}}iv_2\int d^2x \sum_a \left( -i f_a^\dagger
    \partial_2^{\vphantom{\dagger}}f_a^\dagger +if_a\partial_2 f_a\right)\;,
\end{eqnarray*}
with the two velocities given by $v_1 = at$ and $v_2 = a\Delta_0$, and we
have introduced rotated coordinates by $x_1 = {1\over 2}(x + y)$ and
$x_2 = {1\over 2}(x - y)$.  The fermion operators $f_a$ and $f_a^\dagger$ 
are linear combinations of the operators $c_\sigma$ and $c_\sigma^\dagger$
in (\ref{BdG}).
The Hamiltonian $H_0^{(+-)}$ for the nodal point $k_x = - k_y = \pi/2a$ is
obtained from $H_0^{(++)}$ by exchanging $\partial_1 \leftrightarrow
\partial_2$.  The remaining ones, $H_0^{(-+)}$ and $H_0^{(--)}$, are
gotten from the previous two by reversing the overall sign.

We now introduce disorder into the problem.  In the present context,
disorder can be classified into two categories: {\it intra}node
scattering, and {\it inter}node scattering.  We will deal with these
scattering mechanisms separately.  First we take into account the
intranode scattering, and later we will include the internode scattering
as a perturbation.  Such a two-step procedure makes sense if the
``backscattering'' between nodes is weak compared to the ``forward''
scattering within a node.

In the family of symmetry classes of disordered superconductors, class $D$
is distinguished by the absence of any special symmetries such as
time-reversal or spin-rotation invariance.  Thus we are to add disorder of
generic type.  Without much loss, we can restrict the disorder to be {\it
local} (involving no derivatives).  To write down the most general, local,
intranode scattering Hamiltonian, we again single out the node $(++)$ as
an example.  The expression for this Hamiltonian in terms of the operators
$f_a^{\vphantom{\dagger}}, f_a^ \dagger$ reads
  $$
  H_1^{(++)} = {1\over 2}\int d^2x \left( {\textstyle{\sum_{ab}}}
    f_a^\dagger M_{ab}^{\vphantom{\dagger}} f_b^{\vphantom{\dagger}}
    + \Delta f_1^\dagger f_2^\dagger + {\rm h.c.} \right) \;,
  $$
where $\Delta$ now is a complex ``random gap function'', and $M_{ab} 
= \bar M_{ba}$ is a $2\times 2$ ``mass matrix'' fluctuating randomly in 
space.  They are given as certain linear combinations of the
perturbations (random scalar potential, random order parameter, 
{\it etc.})  that are added to the original pure Hamiltonian $H_0$.

Consider now the problem posed by the sum $H^{(++)} = H_0^{(++)} +
H_1^{(++)}$.  This is essentially the problem we solved in Sections
\ref{sec:Qfield}--\ref{sec:DOS}.  To make the correspondence precise,
we choose anisotropic units of length in the $x_1$ and $x_2$ directions
such that $v_1 = v_2 = 1$ (we will restore the proper length units
shortly), and write
\begin{eqnarray*}
  H^{(++)} &=& {1 \over 4} \int d^2x \left( f_a^\dagger \,
    f_a^{\vphantom{\dagger}} \right) \cdot {\cal H}^{(++)}
  \cdot \pmatrix{ f_a\cr f_a^\dagger \cr} + {\rm const} \;, \\
  {\cal H}^{(++)} &=& \pmatrix{
    M_{11} &M_{12} &i\partial_1+\partial_2 &\Delta \cr
    M_{21} &M_{22} &-\Delta & i\partial_1+\partial_2\cr
    i\partial_1-\partial_2 &-\bar\Delta &-M_{11} &-M_{21}\cr
    \bar\Delta &i\partial_1-\partial_2 &-M_{12} &-M_{22}\cr} \;.
\end{eqnarray*}
All information about the second-quantized Hamiltonian $H^{(++)}$ is
encoded in the first-quantized Hamiltonian ${\cal H}^{(++)}$, and we
can equivalently work with the latter instead of the former.  Doing
so, and adopting white-noise disorder
\begin{eqnarray*}
  \left\langle M_{ab}({\bf x}) {\bar M}_{cd}({\bf x}^\prime) \right\rangle
  &=& g \, \delta_{ac}\delta_{bd}\,\delta({\bf x}-{\bf x}^\prime) \;, \\
  \left\langle \Delta({\bf x}) \bar\Delta({\bf x}^\prime) \right\rangle
  &=& g \, \delta({\bf x}-{\bf x}^\prime) \;,
\end{eqnarray*}
with $\langle M_{ab} \rangle = \langle \Delta \rangle = 0$, we arrive
at the $N = 2$ version of the random Hamiltonian treated earlier.
Drawing on the results of Section \ref{sec:gradient}, we can
immediately write down the low-energy effective field theory for this
problem.  (Of course, the anisotropic choice of length units has made
the short-distance cutoff of the field theory anisotropic.  This does
not make a big difference, as the nonlinear sigma model couplings
are essentially cutoff independent.)  On putting the proper units of
length back in place, we get an anisotropic theory ($q = q^{(++)}$),
\begin{eqnarray*}
  S^{(++)} &=& {-1\over 8 \pi}\int d^2x\,{\rm STr}
  \left( {v_1 \over v_2} \, \partial_1 q \, \partial_1 q +
    {v_2 \over v_1} \, \partial_2 q \, \partial_2 q \right) \\
  &+& {1 \over 16} \int d^2x \, \epsilon_{\mu\nu} \,
  {\rm STr}\, q \, \partial_\mu q \, \partial_\nu q \;.
\end{eqnarray*}
Note that the topological term by its nature is ignorant of all length
units and therefore cannot depend on the ratio $v_1/v_2$.  Exactly the
same effective action governs the field $q^{(--)}$ for the node $(--)$.
The actions for the remaining two nodes are obtained by interchanging
$\partial_1 \leftrightarrow \partial_2$, which has the particular
consequence of reversing the sign of the topological coupling.

Finally, we turn on the scattering between nodes, which couples the
four theories together.  Here we can easily guess without any
calculation what is going to happen.  The fields $q^{(st)}$ for the
nodes $(st)$ are the Goldstone modes of a broken orthosymplectic
symmetry.  In the absence of internode scattering, the four sectors
described by the $q^{(st)}$ are decoupled, and the symmetry group
consists of four independent copies of ${\rm OSp}(2n|2n)$.  Scattering
between the nodes reduces the symmetry to the diagonal subgroup, which
acts by the {\it same} factor in each copy.  On these grounds, we
expect that the effect of internode scattering is to ``lock'' the
fields of the four nodes to each other: $q^{(++)} = q^{(--)} =
q^{(+-)} = q^{(-+)} \equiv q$, at large distance scales.  The locked
field $q$ is the Goldstone field due to the remaining diagonal ${\rm
  OSp}(2n|2n)$ symmetry.  The effective action for it will simply be a
sum of actions:
\begin{eqnarray*}
  S_{\rm eff} &=& S^{(++)} + S^{(--)} + S^{(+-)} + S^{(-+)} \\
  &=& - {v_1^2 + v_2^2 \over 4\pi v_1 v_2} \int d^2x \,
  {\rm STr} \, \partial_\mu q \, \partial_\mu q \;.
\end{eqnarray*}
In this final expression the anisotropy of the single-node actions has
cancelled, and $S_{\rm eff}$ is isotropic.  Also, the four topological
terms have added up to zero.  This was to be expected, since the
presence of these terms violates parity, which is a good symmetry of
the superconductor, unless an orientation is induced by a strong
magnetic field or by an order parameter with nonzero chirality (such
as $d_{x^2-y^2}+id_{xy}$).

Thus we have arrived at the nonlinear sigma model for class $D$.  The
model is at weak coupling if the ratio $v_1/v_2$ is large, which is the 
typical situation in experimental systems.  The
coupling $f^{-1} = 4(v_1^2 + v_2^2)/v_1 v_2$ or rather, ${\rm const}
\times T f^{-1}$ where $T$ is the temperature, has an interpretation
as the thermal conductivity of the disordered superconductor
\cite{sfD}.  (A nonzero topological angle $\theta$ would have
corresponded to a thermal Hall conductivity.)

The renormalization group for the weakly coupled theory was worked out
in Section \ref{sec:DOS}.  We found that the theory flows to a
Gaussian fixed point describing a perfect (thermal) metal, and the
local density of states in the thermodynamic limit diverges
logarithmically at $E = 0$.

\section{Free fermions and phase diagram}
\label{sec:phases}

We have shown that a model of $N$ species of Dirac fermions in class
$D$ supports a metallic phase in two dimensions.  Our analysis has put
the existence of that phase on solid ground for $N \gg 1$, or $N = 2$
with large and opposite anisotropies in the velocities.  Here and in
the next section, we wish to go further and address two questions that
emerge from the recent literature \cite{ziegler98,sfD,rg}: (i) Can the
2d metallic phase of class $D$ also be realized in the constrained
parameter space of the fundamental case $N = 1$?  (ii) What is the
location of the free-fermion point relative to the metallic phase?

The answer to the second question is that the metallic phase has the
free-fermion point sitting right on its boundary.  We are going to
demonstrate that this is so, by employing the supersymmetric extension
of nonabelian bosonization developed in Section \ref{sec:gradient}.
As for the first question, we will see that the answer is no, but
there exists a remarkable twist to the story.

We start with the second question.  Recall the Lagrangian $L_2$ in
(\ref{Lag2}), describing $N$ species of massless Dirac fermions plus
$b$-$c$ ghost system, weakly perturbed by the operators $\Phi^{
  (\alpha)}$ with couplings $g_\alpha$ $(\alpha = 1, ...,4)$.  As we
have seen, a single species in the pure limit is equivalent to the
$C|D$ ${\rm WZW}_{k=1}$ model with field $M$ and action functional
$W[M]$.  We now bosonize $L_2$, by exploiting the equivalence for each
species separately:
  $$
  \phi_l^{\rm t} \bar\partial\phi_l^{\vphantom{t}} +
  \bar\phi_l^{\rm t} \partial \bar\phi_l^{\vphantom{t}}
  \rightarrow W[M_l] \qquad (l=1,...,N)\;.
  $$
This scheme can be justified by introducing auxiliary fields $Q$ as in
(\ref{decouple}) to transform $L_2$ into a Lagrangian for $N$
decoupled species.  The perturbations $\Phi^{(\alpha)}$ are then
bosonized using the rules derived in Subsection \ref{sec:nonab}, which
results in
\begin{eqnarray*}
  S_2^\prime &=& \sum_{l=1}^N \left( W[M_l] - {g_1+g_3+g_4 \over
      (2\pi)^2 N} \int d^2x \, {\rm STr}\, J_l \Sigma_3 \bar J_l
    \Sigma_3 \right) \\
  &+& {1\over N} \sum_{k\not= l} \int d^2x\, \Big(
    {g_1 \over \ell^2}\,{\rm STr}\,M_k \Sigma_3 M_l \Sigma_3 +
    {g_2 \over \ell^2}\,{\rm STr}\,M_k^{-1} M_l^{\vphantom{-1}} \\
    &&\hspace{2cm} + {g_3 \over \ell^2}\,({\rm STr}\,M_k \Sigma_3)
    ({\rm STr}\,M_l \Sigma_3) - {g_4 \over (2\pi)^2}
    \,{\rm STr}\,J_k \Sigma_3 \bar J_l \Sigma_3 \Big) \;.
\end{eqnarray*}
Here we isolated the part of the perturbation that acts within a
single species $l$, and bosonized it by reduction to the bosonization
rule for the currents:
  $$
  (\bar\phi_l^{\rm t} \phi_l^{\vphantom{t}})^2 = - {\rm STr}
  (\phi_l^{\vphantom{t}} \phi_l^{\rm t}) (\bar\phi_l^{\vphantom{t}}
  \bar\phi_l^{\rm t}) \to - (2\pi)^{-2} {\rm STr}\, J_l \Sigma_3
  \bar J_l \Sigma_3\;.
  $$
We observe that, since the fields $J_l = M_l^{\vphantom{-1}} \partial
M_l^{-1}$, $\bar J_l = M_l^{-1} \bar\partial M_l^{\vphantom{-1}}$ and
$M_l$ have conformal dimensions $(1,0)$, $(0,1)$, and $(1/2,1/2)$,
respectively, bosonization has preserved the marginality of the
perturbation.  Of course, the couplings will not be truly marginal but
will evolve under renormalization in a way that is determined by the
operator product expansion among the various perturbations.  Given
that the perturbed $C|D$ ${\rm WZW}_{k=1}$ model is a true image of
the original theory, the flow equations are identical to the ones
worked out earlier and given in (\ref{flow}).  Note also that, by
taking the beta functions from the Dirac representation, instead of
recalculating them in the bosonized theory, we avoid the nontrivial
question of how to renormalize the local ${\mathbb Z}_2$ degree of
freedom of the perturbed WZW model.

To go further, we distinguish between cases.  Let us first assume $N >
2$ and take the bare couplings to be $g_1(\ell_0) = g_2(\ell_0)$, and
$g_3(\ell_0) = g_4(\ell_0) = 0$.  Numerical integration of the
one-loop flow equations (\ref{flow}) then shows that the couplings
$g_1, g_2$ grow ($g_1$ more strongly than $g_2$), $g_3$ becomes
nonzero and positive, and $g_4$ moves to negative values but respects
the bound $g_1 + g_4 \ge 0$ dictated by Hermiticity of the random
Hamiltonian.  (For $N = 2$, the flow eventually is attracted to the
line $g_1 + g_4 = g_2 = g_3 = 0$, which has a higher symmetry, namely
that of class $A{\rm II}$; see Section \ref{sec:marginal}.  Thus the
symmetry is dynamically enhanced in that case and the flow, though
starting from a point in class $D$, terminates in class $A{\rm II}$.)
Hence we are facing a relevant perturbation that drives the system
away from the free-fermion point.  We make the reasonable assumption
that the relevant nature of the flow persists beyond the one-loop
approximation used in deriving the flow equations.

The fate of the theory on its way to strong coupling is quite
transparent from the bosonized representation.  First of all, the
effect of the term ${\rm STr}\,M_k^{-1} M_l^{\vphantom{-1}}$ with
large coupling $g_2$ is to ``lock'' the fields, since the function
${\rm STr}\, M_k^{-1} M_l^{\vphantom{-1}}$ has an absolute minimum at
$M_k^{-1} M_l^{\vphantom{-1}} = 1$; recall expression (\ref{minimum}).
Second, in the locked configuration $M_k = M_l \equiv M$ the terms
multiplying $g_1$ and $g_3$ become $\ell^{-2}(N-1)\left( g_1 {\rm STr}
  (M \Sigma_3)^2 + g_3 {\rm STr}^2 (M\Sigma_3) \right)$.  By
parametrizing $M = T {\rm e}^Y T$, with $Y = + \Sigma_3 Y \Sigma_3$
and $\Sigma_3 T \Sigma_3 = T^{-1}$, we can write them as
  $$
  g_1 {\rm STr}(M\Sigma_3)^2 + g_3 {\rm STr}^2 (M\Sigma_3) =
  g_1 {\rm STr}\,{\rm e}^{2Y} + g_3 \left( {\rm STr}\,
    {\rm e}^Y \Sigma_3 \right)^2 \;.
  $$
This is a potential for $Y$ which, from (\ref{minimum}), has an
absolute minimum at $Y = 0$. (Here we again benefit from our careful
construction of the target of the WZW model as a Riemannian symmetric
superspace.)  At strong coupling, we may set $Y = 0$.  Doing so, and
inserting $M_l = T^2$ into the expression for the bosonized action
$S_2^\prime$, we obtain
  $$
  S_2^{\prime\prime} = N W[T^2] + {g_1 + g_3 + N g_4 \over (2\pi)^2}
  \int d^2x\, {\rm STr}\, \partial T^2 \Sigma_3 \bar\partial T^2
  \Sigma_3 \;.
  $$
We finally set $q = T \Sigma_3 T^{-1}$ and identify $q$ with the
field of the nonlinear sigma model.  Then, by a slight extension
of the calculation of Section \ref{sec:gradient}, we find that
$S_2^{\prime\prime}$ reduces to the effective action $S_{\rm eff}[q]$
given in (\ref{NLsM}), with couplings
\begin{equation}
  f = \left( N - ( g_1 + g_3 + N g_4 ) / \pi \right)^{-1} \;,
  \qquad \theta = \pm N \pi \;.
  \label{fval}
\end{equation}

What will be the fate under renormalization of the nonlinear sigma
model so obtained?  Recall that for the choice of bare couplings we
made, $g_1$, $g_2$ and $g_3$ flow to positive values whereas $g_4$
becomes negative.  By linearly combining the basic flow equations
(\ref{flow}) we find
  $$
  \dot g_1 + \dot g_3 + N \dot g_4 = - {(g_1 + g_4)^2 \over N}
  - {(g_3 + g_4)^2 \over N} - ( 1 - {2/N}) g_4^2 - 2 (g_1 + g_2) g_3 \;.
  $$
The right-hand side of this equation is negative, so the combination
$g_1 + g_3 + N g_4$ has a negative rate of change, for all $N > 2$.
(This happens in spite of the fact that the individual couplings
increase in magnitude.)  Hence, the nonlinear sigma model coupling in
(\ref{fval}) is roughly of order $1/N$ and decreases under
renormalization.  It is therefore reasonable to expect that the model
lies in the metallic phase, where the RG flow is attracted by the
Gaussian fixed point $f = 0$.  Our argument then says that the RG flow
takes the Lagrangian $L_2$ with bare couplings $g_1 = g_2$ and $g_3 =
g_4 = 0$ into the metallic phase.  This remains true for arbitrarily
small $g_1 = g_2$, because this coupling is (marginally) relevant.
Thus, no matter how close to the free-fermion point we start, the RG
flow will go to the metallic fixed point.  (Recall that this is not
true for $N = 2$ with equal velocities, in which case the flow is
attracted toward symmetry class $A{\rm II}$.  However, we expect a
finite anisotropy in the velocities to stabilize the flow inside class
$D$.) In other words, for $N > 2$ we are sure of the existence of a
metallic phase, and the free-fermion point sits right on the boundary
of that phase, as claimed.

On the other hand, the situation for $N = 1$ is qualitatively
different.  In that case, the only type of local disorder available in
class $D$ is randomness in the mass.  If we set the statistical
average of the random mass to zero, $m_0 \equiv \langle m(x) \rangle =
0$, there remains only a single coupling $g_M$, which is defined by
(\ref{disorder}).  The bosonized action then reads simply
  $$
  S_2^\prime = W[M] - {g_M \over (2\pi)^2} \int d^2x \,
  {\rm STr}\, J \Sigma_3 \bar J \Sigma_3 \;.
  $$
This is a WZW model perturbed by a current-current interaction.
Computing the beta function directly from the OPEs for the WZW
currents $J$ and $\bar J$, one finds that the perturbation is
marginally irrelevant in the physical range $g_M > 0$, in agreement
with the flow equation (\ref{irrelevant}).  (The perturbation would be
marginally relevant for negative $g_M$.  However, for that sign of the
coupling the $C|D$ WZW model is unstable and does not exist; {\it
  cf.}~the discussion at the very end of Subsection \ref{sec:nonab}.)
Thus, the RG flow is attracted by the WZW fixed point and the system
in this sense is {\it critical}.

The discussion so far assumed $m_0 = 0$.  When a nonzero average mass
is introduced, the bosonized action $S_2^\prime$ acquires an
additional term
  $$
  (i m_0/\ell) \int d^2x\, {\rm STr}\, M\Sigma_3 \;.
  $$
This term is relevant by power counting at the WZW fixed point, and
puts the system off criticality.  Physically speaking, a finite Dirac
mass $m_0$ opens a gap around $E = 0$ in the energy spectrum of the
pure system, which causes localization of all low-energy states due to
disorder (at least as long as $g_M$ is small enough).  Therefore, the
critical line segment $m_0 = 0$, $g_M > 0$ separates two {\it
  insulating} phases.  As is known from \cite{lfsg}, the distinction
between the two phases is of a topological nature.  If one of the two
phases, say $m_0 < 0$, is a plain insulator, the other one $(m_0 > 0)$
can be likened to a quantum Hall fluid, in the sense that the presence
of a boundary gives rise to chiral edge excitations leading to a
quantized Hall-type response.  (In the language of the 2d Ising model
with weakly disordered bond strengths, the two phases correspond to
the paramagnetic and the ferromagnetic phase.)  Thus we have two
insulating phases, and there is no room for a metallic phase, at least
not in the vicinity of the free-fermion point, for $N = 1$.  On
grounds of continuity, the local density of states at zero energy in
these phases is expected to vanish, and we believe the local 
${\mathbb Z}_2$ degree of freedom of the $C|D$ WZW model to play a 
crucial role in reproducing this feature in the field-theoretic
formalism.

\begin{figure}
\epsfysize=6.1cm \epsfbox{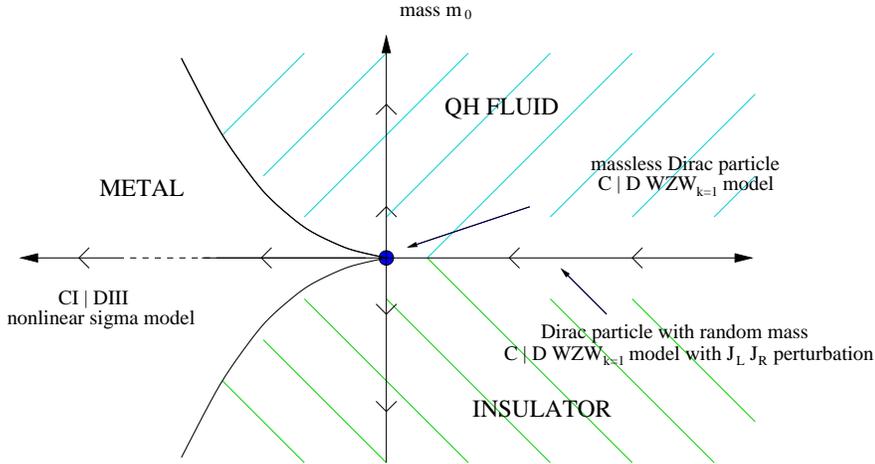}
\caption{\ninerm Schematic structure of the phase diagram close to 
the free-fermion point for class $D$ in two dimensions.  Field-theoretic 
realizations for some parts of the phase diagram are indicated.  The 
coordinate along the horizontal axis is an $N$-dependent function of 
the marginal couplings $g_1,...,g_4$.  For $N = 1$, only the half-axis 
on the right of the free-fermion point is physically realized, with the 
coordinate being $g_M = g_1 + g_3 + g_4 > 0$. The metallic phase 
exists for more than two species, or for $N = 2$ with anisotropy.}
\label{fig:classD}
\end{figure}

By combining the information given, we arrive at Figure
\ref{fig:classD}, which draws a schematic picture of the phase diagram
close to the free-fermion point.  Three phases are seen to meet there:
insulator, quantum Hall fluid, and metal.  (Recall that in the
physical application to superconductors, the terminology ``metal'' and
``insulator'' refers to {\it thermal} transport.)  The two insulating
phases are separated by a critical line, described by the
field-theoretic model of Dirac fermions with random mass, or
alternatively, upon nonabelian bosonization, by the $C|D$ ${\rm
  WZW}_{k = 1}$ model with a current-current interaction.  The flow
along the critical line terminates at the free-fermion point, which
controls the critical behavior across the transition from the
insulator to the quantum Hall fluid.  The metallic phase exists for $N
> 2$ species, or for $N = 2$ with anisotropy in the velocities, but is
absent for $N = 1$.  The field-theoretic model for this phase is the
$C{\rm I}|D{\rm III}$ nonlinear sigma model with couplings $f$ and
$\theta$.  It is expected \cite{sfD} that the Hall response in this
phase is not quantized but varies continuously with the angle
$\theta$.

Two additional remarks are in order.  First, a structurally similar
phase diagram was suggested by Senthil and Fisher \cite{sfD}.  The
main difference is that these authors did not identify the
multicritical point of class $D$ as free fermions and, consequently,
had part of the flow reversed.  Second, recall from Section
\ref{sec:gradient} that the topological term of the $C{\rm I}|D{\rm
  III}$ nonlinear sigma model is {\it trivial} (at least in any
continuum regularization of the field theory) for the case of one
replica ($n = 1$), which contains all information about a single Green
function at any energy $E$, and about a product of two Green functions
at $E = 0$.  It is hard to see how a term which is topologically
trivial could drive a topological phase transition.  This leaves two
options: either the nonlinear sigma model (unlike Pruisken's model for
the integer quantum Hall effect) is not a valid description of the
critical line separating the two insulating phases, or else the
driving agent for the transition is the ${\mathbb Z}_2$ degree of
freedom of the model.  In either case, the sole function of the
topological term is to give rise to the edge current that is expected
to exist for a system with boundary in the metallic and quantum Hall
fluid phases.

\section{Vortices}
\label{sec:vortices}

This is not yet the end of the story.  It has recently been argued by
Read and Green \cite{rg} that {\it vortex} disorder has a significant
effect on the phenomenology of quasiparticle transport and
localization in class $D$.  Without repeating the detailed discussion
of \cite{rg}, we recall the following essential points.

Basic to our study is the spinor field $\psi(x),\psi^\dagger(x)$ of a
Majorana fermion with Hamiltonian
  $$
  H = \int d^2x \left( \psi^\dagger i\partial \psi^\dagger + 
    \psi i\bar\partial \psi + m \psi^\dagger \psi \right) \;, 
  $$
which we here imagine to be the low-energy approximation to a
mean-field Hamiltonian governing the time evolution of the
quasiparticles in a chiral $p$-wave superconductor.  Vortices are
introduced by postulating a change of boundary condition: with
(half-quantum) vortices present, it is no longer true that the
Majorana spinor is a single-valued function of position.  Rather, the
spinor now reverses its sign on circling once around any one of the
vortex singularities.  (We assume the London limit where vortex cores
are point-like objects.)  The phase twist originates from gauging
away the phase of the superconducting order parameter, which winds by
$2\pi$ on going once around a vortex carrying half a magnetic flux
quantum.

Thus the insertion of vortices amounts to the introduction of
square-root singularities into the continuum Majorana field.  Recall
that the one-component Majorana theory in the massless limit ($m = 0$)
is equivalent to the critical 2d Ising model.  Under this equivalence,
the square-root singularities in the Majorana field correspond to spin
fields (or Kadanoff-Ceba disorder fields in a dual picture) placed in
the partition sum of the Ising model.  By using the known \cite{bpz}
conformal dimensions of the spin field at criticality, and
transcribing them to the supersymmetric setting, Read and Green were
able to conclude that the perturbation of the massless free Majorana
theory by randomly placed vortices is strongly relevant.

The question then is: what does the relevant RG flow go to?  Read and
Green \cite{rg} suggested that the flow leads into a metallic phase 
described by a nonlinear sigma model {\it without the local ${\mathbb 
Z}_2$ degree of freedom} which we have identified as being characteristic 
of class $D$.   We have no direct proof of this conjecture at present, as 
we do not know how to incorporate square-root singularities into the 
supersymmetric extension of the nonabelian bosonization scheme.  We 
can, however, give a partial verification as follows.

Throughout our discussion, we are going to assume that the random 
mass $m(x)$ is a smooth function of $x$.  As a warm-up, let us start 
from the homogeneous limit $m(x) = m_0 > 0$ and insert a local 
inhomogeneity with the
shape of a disk-like region ${\cal D}$ where the mass switches to
negative values.  Thus, $m(x)$ is positive outside of ${\cal D}$,
negative inside, and zero along the boundary $\partial {\cal D}$.
The zero-mass contour $\partial{\cal D}$ is a kind of ``inner edge''.
As was nicely explained in \cite{lfsg}, moving a Majorana spinor at
energy $E = 0$ once around $\partial{\cal D}$ reproduces the spinor
with a phase shift of $\pi$.  Hence, Bohr-Sommerfeld quantization
gives a bound state with low energy $E = \pi/L$ (actually, in the
first-quantized formulation, a pair of bound states with energies $\pm
E$), where $L$ is the circumference of the contour.  The localization 
length of the bound state transverse to the contour is inversely 
proportional to the mass $m_0$.

At the critical point $m_0 = 0$, a randomly but smoothly varying mass
function produces a percolating network of zero-mass contours, and
thus a macroscopic number of low-energy states.  To capture the
quantum physics of this random system, we turn to a variant of the
Chalker-Coddington network \cite{cc}, originally designed to model the
quantum percolation transition between plateaus of the integer quantum
Hall effect.  The original model involves random ${\rm U}(1)$ phases
on the links, in addition to scattering at the nodes, of a square
network.  Since we are interested in the limit of zero energy, where
the states propagating along inner edges do not \cite{lfsg} carry any
phases, randomness in the link phases is forbidden in the present
context.  Thus we are led to study a network model without any random
phases, just randomness in the probability for scattering to the right
($p_R$) or left ($p_L$).  At the symmetric point of the model, where
$p_L = p_R = 1/2$ on average, the continuum limit is known \cite{hc}
to be a Dirac fermion (or, equivalently, two copies of a Majorana
fermion) with random mass and $m_0 = 0$.  Thus we have come full
circle and are back to our starting point.

Consider now the effect of adding vortices to the system.  As we
recall, going once around a vortex in the continuum formulation twists
the phase of the spinor wavefunction by minus one.  For the case of an
isolated zero-mass contour enclosing one vortex, the phase shift by
$\pi$ has the important consequence of giving rise to a single {\it
  zero-energy bound state} or fermionic zero mode \cite{volovik}.  In
what follows, we shall be concerned with the effect of vortices on the
critical system.  This is captured in the network model by introducing
``frustrated'' plaquettes, by which we mean that the product of the
link phase factors along any loop encircling such a plaquette equals
minus one.  Thus, randomly placed vortices in the continuum theory
translate into randomly placed frustrated plaquettes in the network
model.  To create an isolated frustrated plaquette, we insert a
semi-infinite string of minus signs on links, terminating at the
plaquette.  To create a large amount of frustration, we insert many
strings.  The maximal amount of disorder, corresponding to a high
density of vortices, is realized by taking the link phases to be
independent and identically distributed random variables drawn from
the uniform distribution on ${\mathbb Z}_2=\{\pm 1\}$, which is the
same as the orthogonal group in 1 dimension, ${\rm O}(1)$. In this
limit we arrive at the Chalker-Coddington network model with a single 
channel per link and ${\mathbb Z}_2$ invariant link disorder.

The simplest choice is to take the scattering probabilities to be
nonrandom.  Then, by the color-flavor transformation developed in
\cite{network} for ${\rm U}(N)$ (which readily extends \cite{beuchelt}
to ${\rm O}(N)$ for all $N$ including $N = 1$) or, alternatively, by
Read's second-quantized setup \cite{grs}, the ${\mathbb Z}_2$ phase
invariant network model at its left-right symmetric point readily maps
on the $C{\rm I}|D{\rm III}$ nonlinear sigma model with action
(\ref{NLsM}), and couplings $f = 2/\pi$, $\theta = \pi$.  However,
there is one important difference from before, which is that the
target space now has just {\it one} connected component.  This 
can be understood by anticipating from Section \ref{sec:rmtDB} 
that $s \in {\mathbb Z}_2$ acts on the identity
component by the trivial representation $|s| = 1$, and on the other
component by the faithful representation $s = \pm 1$.  The latter is
wiped out by averaging over the uniform distribution on ${\mathbb
  Z}_2$.  (It reappears when the distribution is made nonuniform.)
Since the free-fermion point formally corresponds (by saddle-point
approximation in the limit of small $g_M$ for $N = 1$; see Section
\ref{sec:gradient}) to the nonlinear sigma model with $f = 1$, and
since the absence of a disconnected component of the target space does
not change the perturbative RG beta function, the model with a
weakened coupling $f = 2/\pi < 1$ is expected to flow to the metallic
fixed point $f = 0$.  This result is compatible with the expectation
of Read and Green that the addition of vortex disorder at the
free-fermion point results in metallic behavior.  By computing the
functional integral of the nonlinear sigma model in zero-mode
approximation, we find that the density of states for a finite 
network in the ergodic regime is
\begin{equation}
  \rho(E) = \nu + {\textstyle{1\over 2}} \delta(E) \;.
  \label{dosBD}
\end{equation}
Note that this answer {\it differs} from the ergodic limit of the
density of states for quasiparticles in class $D$; {\it cf.}~Section
\ref{sec:DOS} and equation (\ref{dosD}) below.

Thus, vortices change the low-energy density of states in the metallic
phase.  The change is relatively minor there, being visible only on
the microscopic scale of the mean level spacing $\nu^{-1}$.  In the
insulating phases, however, vortices are expected to have a much more
pronounced effect.  We have seen that, when vortices are absent, the
local density of states at zero energy vanishes in that case.  In
contrast, the insertion of a finite density of vortices gives rise, by
the above reasoning (for a slowly varying random-mass function), to an
extensive number of quasiparticle states at very low energy.  In an
insulating phase these do not communicate over large distances, and
therefore they do not repel, leaving a finite density of states at $E
= 0$.  In the field-theoretic setting, we speculate that this comes
about by the mechanism we identified above: {\it vortices suppress the
  local ${\mathbb Z}_2$ degree of freedom} of the nonlinear sigma
model.

We believe these differences in local observables and their
field-theoretic manifestation to be sufficiently significant to
warrant a refinement in vocabulary.  What we propose to say is that
Majorana fermions subject to vortex disorder belong to a generic
(hybrid) class $BD$.  This nomenclature is motivated by the following
considerations.

\subsection{Random-matrix limits}
\label{sec:rmtDB}

The fundamental point we are going to make, here, is that the Lie
algebraic structure underlying the Majorana fermion provides us with
more than one kind of universal level statistics, and this suggests a
refinement of the symmetry classification scheme.  We begin by
recalling from \cite{az} that the basic random-matrix model for class
$D$ is the Gaussian ensemble over $D_N = {\rm so}(2N)$, the Lie
algebra of the orthogonal group in {\it even} dimension.  A Hermitian
matrix drawn at random from $iD_N$ has $N$ pairs of eigenvalues $\pm
E$, with the universal large-$N$ limit of the density of states being
\begin{equation}
  \rho_D(E) = \nu + {\sin(2\pi\nu E) \over 2\pi E} \;.
\label{dosD}
\end{equation}
Note that this function is positive and smooth at $E = 0$. 

Another Lie algebra, regarded as mathematically distinct from ${\rm
  so}(2N)$ (as the root systems differ in structure) is that of
the orthogonal group in {\it odd} dimension, denoted by $B_N = {\rm
  so}(2N+1)$ in Cartan's notation.  The distinction becomes expecially
evident in the level statistics: the density of states of the Gaussian
ensemble over $i B_N$ has the large-$N$ limit
\begin{equation}
  \rho_B(E) = \nu - {\sin(2\pi\nu E) \over 2\pi E} + \delta(E) \;.
\label{dosB}
\end{equation}
Here we see a singular term $\delta(E)$, reflecting the fact that an
antisymmetric matrix in odd dimension has one, and generically only
one, eigenvalue at zero.  (To understand this fact by way of example,
recall that every proper rotation of Euclidian 3-space has one
invariant vector, namely the axis of rotation.)  Repulsion from that
level is the cause of the quadratic law $\rho_B(E) \sim E^2$ near $E =
0$. We mention in passing that the above law has also been found, in
recent mathematical work \cite{ks}, to be the universal limit of the
density of zeros of suitably chosen families of algebraic
$L$-functions.

What we have reviewed and illustrated, then, is the fact that the
large family of ``orthogonal'' Lie algebras splits into two classes,
conventionally denoted by $D$ and $B$.  The latter class was mentioned
in \cite{az} but not pursued there for want of a good physical
example.  Recently, it has been proposed \cite{ivanov} that the
universal random-matrix limit of the density of bound states of a
disordered vortex in a chiral $p$-wave superconductor, is given by
class $B$.  In earlier work \cite{sfk,bcsz1,bhl}, the same limit for a
conventional $s$-wave superconductor had been shown to be governed by
class $D$.

What is the relevance of all this to systems in the thermodynamic
limit, when an extensive number of vortices is present?  To answer
that question, notice first of all that the ergodic limit
(\ref{dosBD}) of the density of states for the ${\mathbb Z}_2$ phase
invariant network model coincides with the {\it arithmetic mean} of
the random-matrix answers for the classes $B$ and $D$.  This is
already a first indication that the network model should {\it not} be
assigned to the symmetry class $D$ (at least not in a narrow sense),
but is better placed in a hybrid class, which we propose to call $BD$.
We are going to elaborate below.

For the moment, we shall explain why the coincidence of (\ref{dosBD})
with the arithmetic mean ${1 \over 2}(\rho_D + \rho_B)$ is not an
accident.  Let us interpret \cite{km} the network model as a quantum
dynamical system evolving in time by discrete steps of size $\Delta
t$.  The discrete time-evolution operator $U = \exp -iH\Delta t$ is a
product of two unitary operators, one of which is diagonal (in the
link basis) containing all the random ${\mathbb Z}_2$ factors
associated with the links, while the other one encodes the random
tunneling at the nodes.  The eigenvalues of $U$ are written ${\rm
  e}^{-iE\Delta t}$, where $E$ is called the quasi-energy.  Let the
network now be finite, with an even number $2N$ of links.  Then $U$,
being a unitary matrix with real entries, is an element of the
orthogonal group ${\rm O}(2N)$.  Note that for the model with
${\mathbb Z}_2$ phase invariant disorder, the probabilities for $U$ 
to lie in the proper and improper parts of ${\rm O}(2N)$ are equal.

Thus the time-evolution operator of the network model is a (sparse)
random matrix on ${\rm O}(2N)$, and in order to understand the ergodic
low-energy limit of the density of quasi-energies $E$, one needs to
work out the random-matrix level statistics for that group.  With some
linear algebra, one can show that the level statistics for the two
Haar ensembles on the proper and improper components of ${\rm O}(2N)$
{\it coincide exactly} with those of the classes $D$ and $B$, respectively.
(A noteworthy feature here is that, since the nonreal eigenvalues of
an orthogonal matrix come in complex conjugate pairs, every $U$ in the
improper component of ${\rm O}(2N)$, where ${\rm Det}U = -1$, must
have an eigenvalue $-1$, which then is accompanied by a corresponding
eigenvalue $+1$, giving the delta function in (\ref{dosB}).)  The
general answer for the universal limit of the density of states is
some linear combination of $\rho_D$ and $\rho_B$ as given by
(\ref{dosD}) and (\ref{dosB}).  When the weights assigned to the
proper and improper components of ${\rm O} (2N)$ are equal, the answer
equals the arithmetic mean, which reproduces the result
(\ref{dosBD}), obtained from the nonlinear sigma model.

Of course, the disordered quasiparticle systems we are interested in
are typically very far from the random-matrix limit.  One might
therefore think that a nomenclature borrowed from that limit is too 
crude to capture the general situation.  We insist, however, that the
terminology we propose is fully supported from field theory, by the
following argument.

The key is to understand how the expressions (\ref{dosD},\ref{dosB})
arise from the zero-dimensional limit of the field-theoretic
representation by a nonlinear sigma model.  As we recall, the target
space of the field theory for class $D$ has two connected components,
and the result (\ref{dosD}) comes from summing over both of them with
equal Boltzmann weights.  By inspection, one finds that the expression
for class $B$ results from a {\it twisted} target, where the Boltzmann
weight of the component not containing the identity element carries a
minus sign.  It is useful to let the group ${\mathbb Z}_2$ act on the 
target space, so that twisting the Boltzmann weights is the same as 
acting with the nontrivial element of ${\mathbb Z}_2$.
Now recall that the insertion of vortices amounts to creating
frustrated plaquettes in the network model.  Frustration of
plaquettes, in turn, amounts to the presence of (strings of) links
with minus signs.  Any odd number of these will transfer the
time-evolution operator $U$ from the proper to the improper part of
the orthogonal group, and vice versa.  The crucial proposition we are
driving at is that ${\mathbb Z}_2$ vortex disorder acts as ${\mathbb
  Z}_2$ sign disorder on the target space of the nonlinear sigma model.  
(Indeed, as we saw, going from $D$ to $B$ in the random-matrix limit
is like switching between the proper and improper components of ${\rm
  O}(2N)$.) This proposition, albeit motivated by random-matrix
considerations, does not depend on the random-matrix limit and we are
therefore confident to predict its general validity.  Thus, we expect
that a square-root singularity (due to a half-quantum vortex) at a 
point $p$ in the continuum Majorana field acts as the nontrivial element
of ${\mathbb Z}_2$ at $p$ on the target space of the $C{\rm I}|D{\rm III}$
nonlinear sigma model (whenever that model is a valid description);
and, hence, {\it averaging over randomly placed vortices suppresses
  the second target space component of that model}.

This, ultimately, is the field-theoretic reason why vortices change the 
physics of disordered Majorana fermions, and why we believe that the 
subclass $D$ (without vortices) ought to be kept distinct from the more 
generic hybrid class $BD$.

\section{Open questions}

An extensive summary of our main results was already given in the
introduction and will not be repeated here.  Instead, we content
ourselves with pointing out a few open questions we find interesting.

i) It has been conjectured \cite{ivanov} that the spectral statistics
of the bound states of an isolated disordered vortex in a chiral
$p$-wave superconductor, coincides with the random-matrix statistics
of class $B$ in the universal limit.  It would be desirable to try and
prove that conjecture, by adapting the treatment of \cite{bcsz1} to
derive the nonlinear sigma model for class $B$ from a microscopic
model.

ii) To elucidate the effect of vortex disorder on a system close to
the free-fermion point, it would be useful to know how to correctly
transform the vortex (or square-root) singularities in the Majorana
field to the bosonized representation.  Since the insertion of a
square-root singularity changes the Neveu-Schwartz fermions of the
radial quantization scheme into Ramond fermions, the question is how
to modify the $C|D$ ${\rm WZW}_{k=1}$ model to account for a
degenerate (Ramond) vacuum.  

iii) Although our work sheds much light on the 2d metallic phases of
the classes $D$ and $BD$, the insulating phases remain poorly
understood.  There, we expect domain walls in the ${\mathbb Z}_2$
degree of freedom of the $C{\rm I}|D{\rm III}$ nonlinear sigma model
to be important in determining the low-energy physics.

iv) According to \cite{cf}, the 2d random-bond Ising model (with
fluctuating signs, not bond strengths) can be cast in the form of a
Chalker-Coddington network model with frustration (or vortices) on
adjacent plaquettes.  It would be interesting to understand better how
this model fits into the expanded landscape of class $BD$ versus $D$.
In particular, the nature of the multicritical Nishimori point remains to
be clarified. \medskip

{\bf Acknowledgment}.  M.R.Z. acknowledges J. Chalker for drawing his
attention to the paper of Read and Green, and M. Feigel'man for a helpful
remark on fermionic zero modes in chiral $p$-wave superconductors.
D.S. thanks E. Fradkin for questioning the validity of the saddle-point
approximation for $N = 1$.  She also thanks D. Bernard for useful
discussions.

\end{document}